\documentclass [superscriptaddress, prb, twocolumn,bibliography]{revtex4-2}
\usepackage{graphicx,dcolumn,bm,amssymb,amsmath,hyperref,epsf,xcolor,float}
\usepackage{hyphenat}
\def\ET{$\alpha$-(BEDT\--TTF)$_2$I$_3$}
\def\STF{$\alpha$-(BEDT\--STF)$_2$I$_3$}
\def\BETS{$\alpha$-(BEDT\--TSF)$_2$I$_3$}

\DeclareRobustCommand{\ETI}{%
  \ensuremath{\alpha}\nobreakdash-(ET)\textsubscript{2}I\textsubscript{3}%
}

\DeclareRobustCommand{\STFI}{%
  \ensuremath{\alpha}\nobreakdash-(STF)\textsubscript{2}I\textsubscript{3}%
}
\DeclareRobustCommand{\BETSI}{%
  \ensuremath{\alpha}\nobreakdash-(BETS)\textsubscript{2}I\textsubscript{3}%
}

\def\cm{cm$^{-1}$}
\def\TMIT{$T_{\rm MIT}$}

\usepackage{booktabs}

\begin{document}

\title{Charge-localization-driven metal-insulator phase transition \\ in layered molecular conductors}

\author{Savita Priya}
\affiliation{1. Physikalisches Institut, Universit\"at Stuttgart, Pfaffenwaldring 57,
70569 Stuttgart, Germany}
\author{Maxim Wenzel}
\affiliation{1. Physikalisches Institut, Universit\"at Stuttgart, Pfaffenwaldring 57,
70569 Stuttgart, Germany}
\author{Olga Iakutkina}
\affiliation{1. Physikalisches Institut, Universit\"at Stuttgart, Pfaffenwaldring 57,
70569 Stuttgart, Germany}
\author{Marvin Schmidt}
\affiliation{1. Physikalisches Institut, Universit\"at Stuttgart, Pfaffenwaldring 57,
70569 Stuttgart, Germany}
\author{Christian Prange}
\affiliation{1. Physikalisches Institut, Universit\"at Stuttgart, Pfaffenwaldring 57,
70569 Stuttgart, Germany}
\author{\\Dieter Schweitzer}
\affiliation{1. Physikalisches Institut, Universit\"at Stuttgart, Pfaffenwaldring 57,
70569 Stuttgart, Germany}
\author{Yohei Saito}
\affiliation{1. Physikalisches Institut, Universit\"at Stuttgart, Pfaffenwaldring 57,
70569 Stuttgart, Germany}
\author{Reizo Kato}
\affiliation{RIKEN, 2-1, Hirosawa, Wako-shi, Saitama 351-0198, Japan}
\author{Koichi Hiraki}
\affiliation{Center for Integrated Science and Humanities, Fukushima Medical University, Fukushima 960-1295, Japan}
\author{Martin Dressel}
\affiliation{1. Physikalisches Institut, Universit\"at Stuttgart, Pfaffenwaldring 57,
70569 Stuttgart, Germany}

\date{\today}

\begin{abstract}
The organic conductor $\alpha$-(BEDT-TTF)$_2$I$_3$ provides the prime example of a charge-order-driven metal-insulator transition. Restricted chemical substitution of S atoms by Se in the constituent molecules allows us to modify the electronic properties. This not only decreases the transition temperature but, in addition, alters the phase transition mechanism, resulting in the ground state deviating from the charge-ordered insulator state of the parent compound. Employing infrared optical spectroscopy, we investigate changes in the charge dynamics. Furthermore, we demonstrate the absence of charge ordering in the Se-substituted materials and suggest that the phase transition is instead driven by the localization of the itinerant charge carriers due to strong electron-phonon interactions.
\end{abstract}

\maketitle

\section{Introduction}

Metal-insulator transitions (MIT), observed in a wide range of solids, were originally explained by classical band theory. However, more recent research has challenged and extended this view, highlighting the roles of disorder, correlations, and frustration, along with the complex physics complementing these mechanisms \cite{dobrosavljevic2012conductor, mott2004metal, imada1998metal, ung1994metal, altshuler2001metal}. Molecular conductors based on BEDT-TTF charge-transfer salts [where BEDT-TTF stands for bis\-(ethylene\-di\-thio)tetra\-thia\-fulvalene] serve as important models to unveil the underlying electronic properties of solids and have been under scrutiny for studying the MIT, for many years. In these quasi-two-dimensional systems, the effects of frustration and correlations can be readily tuned either by hydrostatic pressure, by chemical pressure in the anion layer, or the targeted variations of the constituent molecules \cite{jepsen2006metal, kanoda1997electron, abrahams1996metal, mori1999structural, kino1998phase, mori1998structural, miyagawa2004nmr, riedl2022ingredients, tanatar2002pressure}.

Among the class of organic conductors, \ET{} is one of the best-studied compounds. Its salient feature is the MIT at $T_{\rm MIT}=135$~K, which initially could not be explained by structural modulation, thermal effects, and electronic correlations \cite{bender1984synthesis}. The insulating phase was later confirmed to be horizontal-stripe charge-ordered by X-ray synchrotron measurements, which is in accordance with the correlation-driven charge-ordered state predicted by the extended Hubbard model \cite{seo2000charge, kakiuchi2007charge}; by now, numerous spectroscopic methods have formed a consistent picture \cite{Takano01,Takahashi06,Tomic2015Ferro}.
Further phenomena subsequently discovered, such as the phase coexistence in the vicinity of the transition and a putative high-pressure massless Dirac fermionic state,  keep scientists puzzled even after 40 years of continuous research \cite{pustogow2018internal, piechon2013dirac, uykur2019optical}. The studies have eventually extended to iso-structural compounds, namely \STF{} and \BETS, initially to explore states similar to \ET, as they also exhibit MIT. Nowadays, the research on these compounds have been augmented to investigate the alterations in the electronic ground states by chemical modification. Albeit the literature on the phase transitions in \STF{} and \BETS{} is vast, it yet remains indecisive on the driving force and the exact mechanism for MIT \cite{kondo2009crystal, hiraki2011local, tsuchiya2023ultrafast, tsuchiya2023effects}.

\begin{figure}
    \centering
        \includegraphics[width=0.85\linewidth]{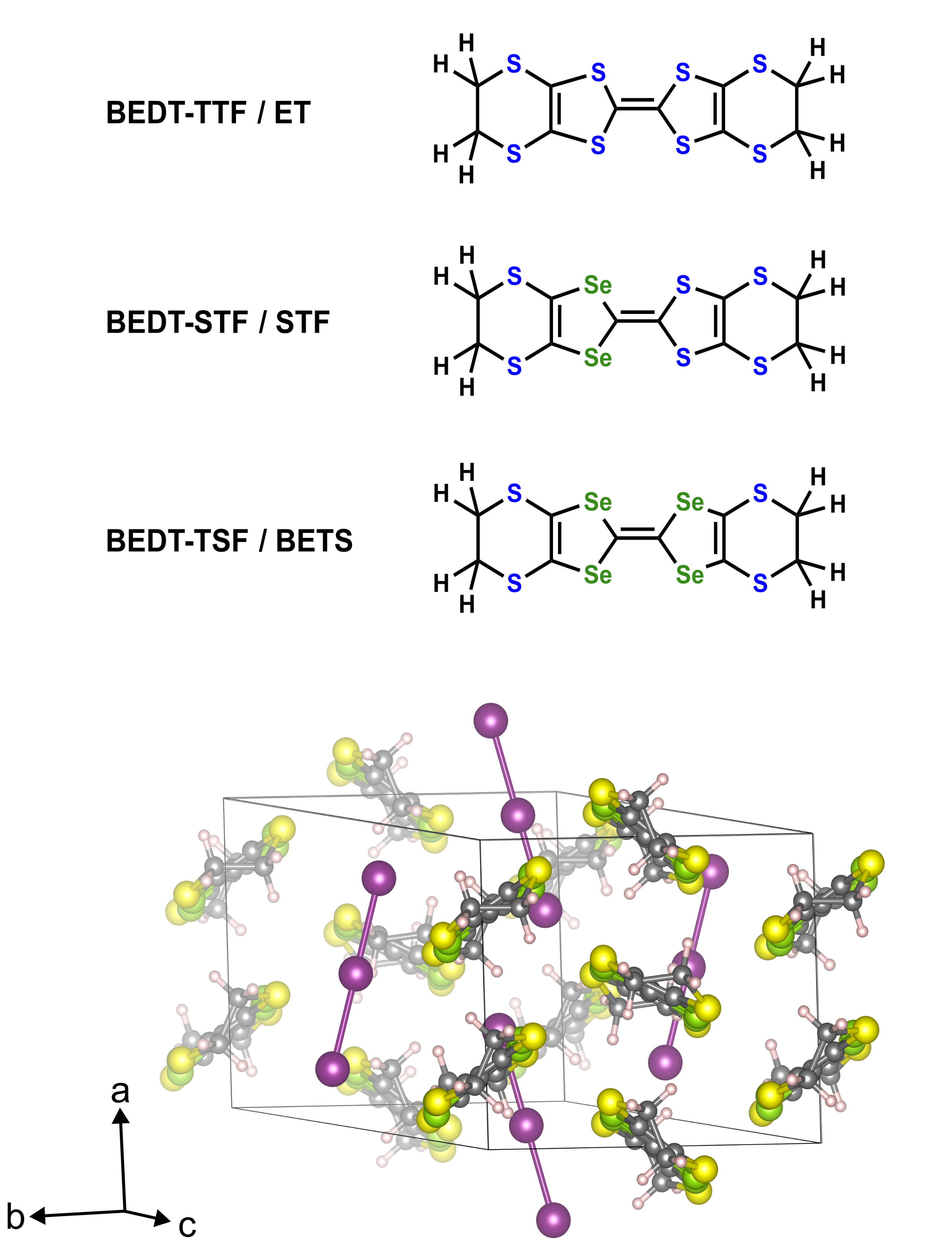}
        \caption{Organic framework of BEDT-TTF, BEDT-STF and BEDT-TSF formed on substitution of S atoms by Se. The crystal structure (bottom) depicts the herringbone arrangement of these organic entities forming the $\alpha$-pattern of the conducting organic layer, sandwiched between insulating I$_{3}^{-}$ layers.}
    \label{fig:Fig1}
\end{figure}

In our current broadband infrared spectroscopy study, we investigate the effects of this chemical substitution in the conducting layer of \ET\ (hereafter \ETI), forming \STF\ and \BETS\ (hereafter \STFI\ and \BETSI, respectively; see Fig.~\ref{fig:Fig1} for illustration), and its influence on the metal-insulator phase transition. By employing a detailed analysis of the temperature-dependent electronic response and the molecular vibrations of the organic layers, we demonstrate that the substitution not only alters the phase transition mechanism and the insulating phase, but also significantly affects the electronic ground state of the metallic phase itself. Unlike the charge-ordered state in \ETI, the MITs in the Se-substituted compounds occur without major structural changes or strong modifications in band structures for \BETSI.

\section{Materials and methods}

\ETI, \STFI{} and \BETSI{} single crystals were synthesized by standard electrochemical synthesis method described in Refs.~\onlinecite{bender1984synthesis, naito1997electrical, kato1991synthesis}. The crystals are flake-like with a flat two-dimensional surface, having typical dimensions of $1.2 \times 0.8~{\rm mm}^{2}$ and a thickness of 20-50~$\mu$m.
Transport measurements were performed by a standard four-probe setup, making contacts with gold wires and carbon paste on the specimen, with the four contacts placed on a straight line in the flat conducting plane. The data were recorded while warming up to avoid jumps in the values due to micro-cracks.

The optical response was probed by optical reflectivity measurements on as-grown samples, along $E \parallel a$, $E \parallel b$ (within the flat surface) and $E \parallel c$ (normal to the plane). For the frequency range of 600-18000~\cm, a Bruker Vertex 80v Fourier-Transform Infrared (FT-IR) spectrometer equipped with a Hyperion microscope and a CryoVac Konti Micro helium-flow cryostat were deployed.  In the lower frequency region of 80-600~\cm, the reflectivity was measured using a Bruker IFS113v FT-IR spectrometer with a custom-built cryostat, equipped with an \textit{in situ} gold overcoating unit. For the metallic temperature range, the data below 80~\cm were extrapolated using the Hagen-Rubens relation, while for the insulating range, they were extrapolated with a constant value. Above 18000~\cm, the data were extrapolated by X-ray scattering functions \cite{tanner2015use}. For $E \parallel c$, the response remains insulating in the entire temperature range; hence we extrapolated assuming a constant behavior below 700~\cm.
The standard Kramers–Kronig relation was employed to calculate the optical conductivity $\sigma_1(\omega)$ from the reflectivity data $R(\omega)$, as described in Ref.~\onlinecite{dressel2002electrodynamics}.

The obtained optical conductivity is fitted by the Drude-Lorentz model to resolve different contributions, also including Fano fit-functions for the asymmetric and sharp molecular vibrations (See Supplemental Material \cite{SM} for more details) \cite{dressel2002electrodynamics, damascelli1997infrared}.

Density-functional-theory (DFT) calculations of the band structure and optical conductivity were performed with Quantum Espresso \cite{giannozzi2009quantum, giannozzi2017advanced} following Ref.~\onlinecite{ohki2023gap}. Self-consistent calculations were converged on the $5 \times 5 \times 3$ $k$-mesh while the optical conductivity was calculated on a denser $k$-mesh with up to $8 \times 8 \times 8$ points.

\section{Results and analysis}

In the following Sections, we will extensively examine the possibilities of different conventional mechanisms (such as thermal activation and disorder) as well as correlation-driven phenomena (charge-ordering), while also exploring possible hallmarks of novel mechanisms in driving the MIT in \STFI\ and \BETSI. These analyses are carried out in direct comparison with the electronic response of the parent compound, \ETI, to elucidate the consequences of selective Se substitution.

\subsection{Transport measurements}

\begin{figure} [h]
    \centering
        \includegraphics[width=1\linewidth]{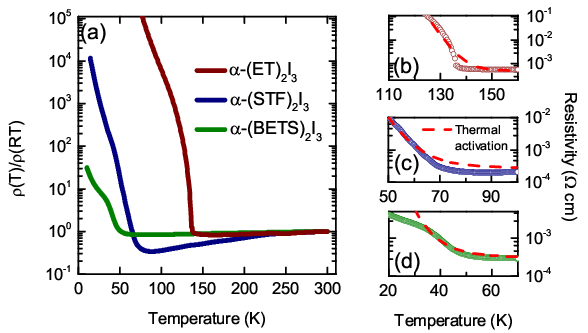}
        \caption{(a) Temperature-dependent resistivity of  \ETI, \STFI, and \BETSI, measured along the conducting $ab$-plane, illustrating the MIT transition by the drastic increase at 135 K for \ETI, 66 K for \STFI, and 50 K for \BETSI; (b-d) Fits by Arrhenius equation cannot capture the abrupt upturn in resistivity at MIT for (b) \ETI, (c) \STFI, and (d) \BETSI.}
    \label{fig:Fig2}
\end{figure}

\begin{figure*}
    \centering
        \includegraphics[width=0.75\linewidth]{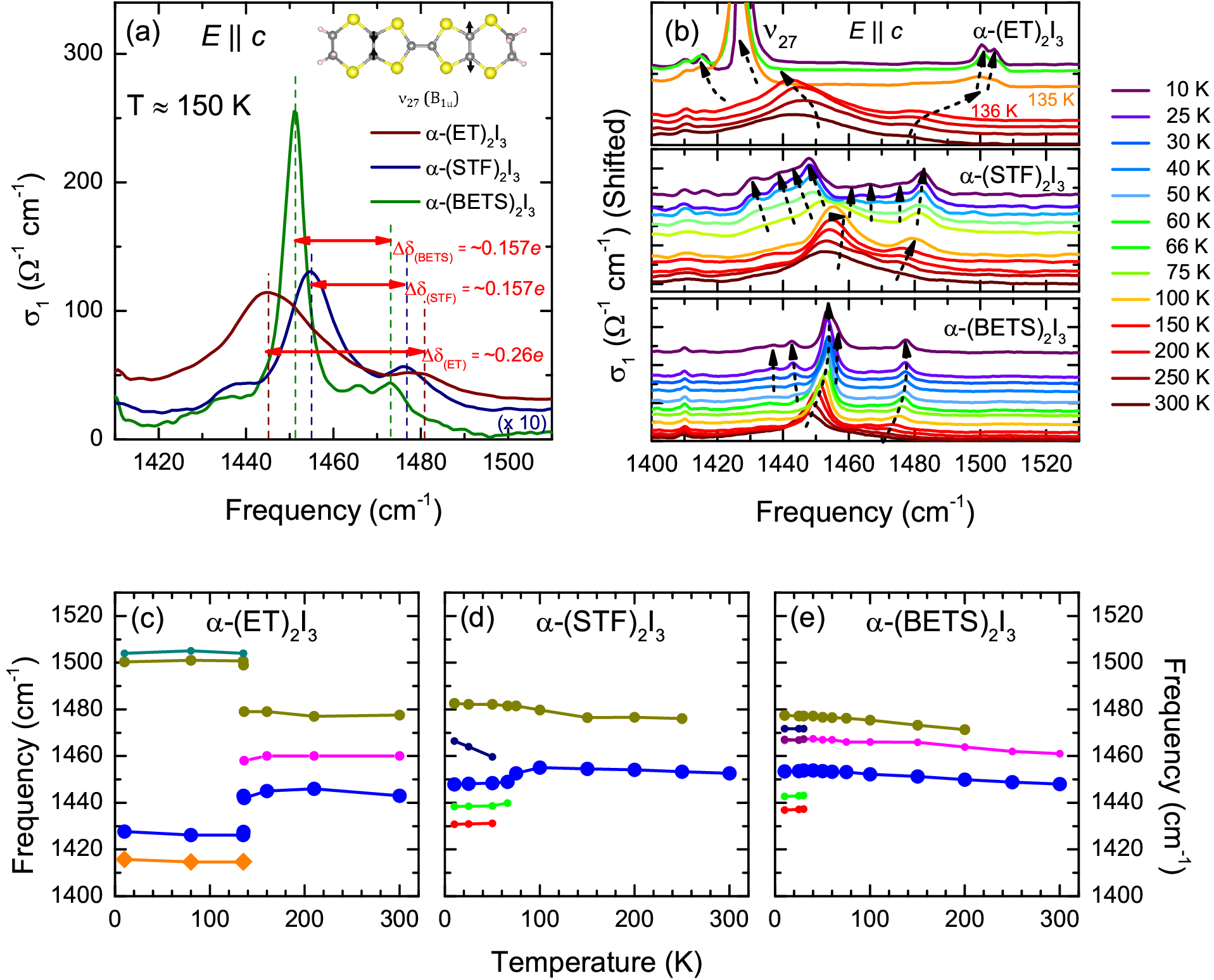}
        \caption{(a)~Out-of-plane ($E \parallel c$) optical response in the range of the infrared-active charge-sensitive vibrational mode $\nu_{27}$ (depicted in the inset) for the three compounds under scrutiny here. Already above \TMIT, multiple peaks indicate charge disproportionation; compared to \ETI\ it is lower for \STFI\ and \BETSI. (b)~Evolution of the spectra upon cooling from $T=300$ to 10~K. \ETI\ exhibits an abrupt splitting, \STFI\ demonstrates unusual broadening and eventual separation of multiple peaks, while in the case of \BETSI, the $\nu_{27}$ feature does not exhibit a prominent splitting but develops several small peaks and a shoulder peak near phase transition. (c-e)~Temperature dependence of the different peak positions extracted from the spectra shown in panel~(b). The data for \ETI{} are reproduced from Ref.~\onlinecite{ivek2011electrodynamic}.}
    \label{fig:Fig3}
\end{figure*}

Fig.~\ref{fig:Fig2} displays the dc-resistivity of the three $\alpha$-phase compounds. The temperature-dependent resistivity is in accordance with previous results, confirming the MIT in \ETI, \STFI\ and \BETSI\ at 135, 66, and 50~K, respectively \cite{inokuchi1995electrical}. At first glance, the significantly lower phase transition temperature for both Se-substituted compounds resembles the suppression of the MIT in \ETI{} by hydrostatic pressure \cite{beyer2016pressure}. Previous studies, however, indicated that chemical substitution decreases the transverse inter-molecular interactions, leading to more complex effects \cite{inokuchi1995electrical, kondo2009crystal, hiraki2011local, tsuchiya2023ultrafast, tsuchiya2023effects}.

The resistivity was further analyzed to examine whether the phase transition in \STFI\ and \BETSI\ is driven by thermal activation similar to narrow-gap semiconductors (see Fig.~\ref{fig:Fig2} (b-d)) or by disorder (following variable-range hopping mechanism) \cite{SM, baranovski2006charge, mott2012electronic}. We conclude that the Arrhenius law cannot describe the resistivity of \STFI\ and \BETSI\ as it does not reproduce the abrupt and sharp phase transition. Additionally, the variation of the slope within the insulating state around 50 and 30~K, for \STFI\ and \BETSI, respectively, points toward a more complex driving force and rules out a simple disorder-driven phase transition in \STFI\ and \BETSI. More details are presented in the Supplemental Material \cite{SM}.

\begin{figure*}
	\centering
    	\includegraphics[width=0.9\textwidth]{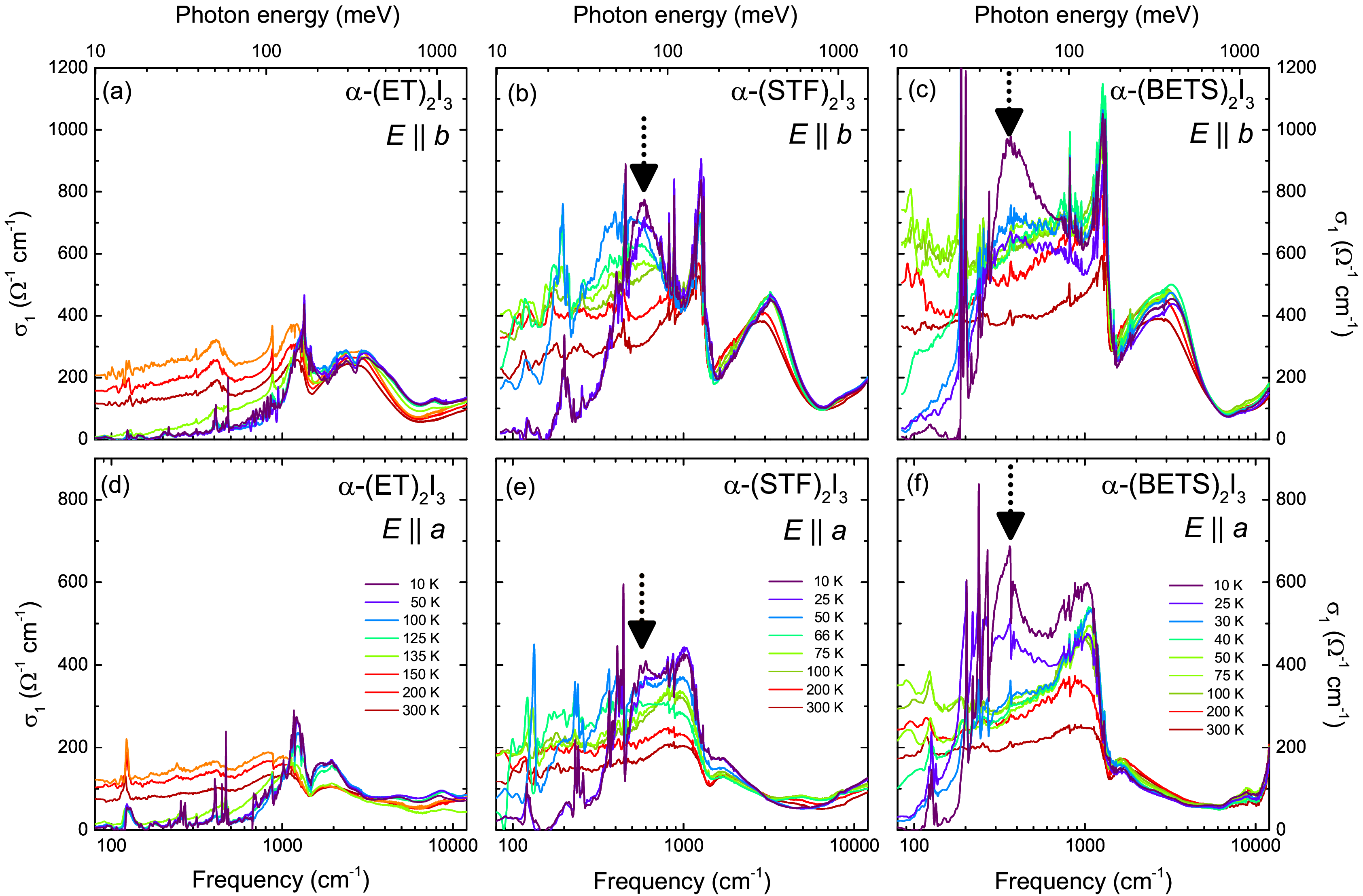}
    	\caption{Temperature-dependent optical conductivity in the range of 10 to 300~K for two different in-plane polarization axes ($E \parallel a$ and $E \parallel b$) for (a,d) \ETI, (b,e) \STFI, and (c,f) \BETSI. The arrows indicate the  center of the absorption peak in the low-energy region for \STFI\ and \BETSI.}
    \label{fig:Fig4}
\end{figure*}

\subsection{Charge-sensitive molecular vibrations: Charge ordering?}
Vibrational spectroscopy by Raman and infrared experiments has been established as primary method for detecting charge disproportionation in molecular charge-transfer salts \cite{dressel2004optical, Yakushi2012}.
The infrared-active charge-sensitive mode $\nu_{27}$ (asymmetric stretching of the C=C bonds) is clearly observed for $E \parallel c$ and is well isolated from other molecular vibrations in terms of energy. A thorough analysis of the molecular vibrations of \ETI\ \cite{ivek2011electrodynamic} identifies multiple peaks for $\nu_{27}$ when the system is metallic (representing different types of BEDT-TTF molecules in two distinct stacks);
the abrupt splitting of the most intense peaks and significant shift of the peaks
mark the transition to a charge-ordered insulating state, depicting the charge disparity of $\delta \approx 0.6e$ in the horizontal stripes.

The spectra of molecular vibrations in the region of the $\nu_{27}$ mode for \ETI\ and Se-substituted compounds are shown in Fig.~\ref{fig:Fig3} \footnote{The notations for molecular vibrations like $\nu_{27}$,  which are known for BEDT-TTF \cite{dressel2004optical}, are analogously used for BEDT-STF and BEDT-TSF in our discussion.}.
At temperatures well above the phase transitions, both \STFI\ and \BETSI\ possess 
multiple peaks [Fig.~\ref{fig:Fig3}(a)], which are explained by the occurrence of dissimilar stacks in the crystal structure as in the case of \ETI. By calculating the charge disparity  above the phase transition based on the two most intense peaks, utilizing \cite{yamamoto2005examination}:
\[
\nu_{27}(\delta) = 1398 \,\text{cm}^{-1} + 140(1 - \delta) \,\text{cm}^{-1} \quad ,
\]
we can estimate the charge separation of approximately  $0.15e$ in \STFI\ and \BETSI, which is significantly smaller than in \ETI.

While the $\nu_{27}$ mode in \STFI\ shows an abrupt splitting into multiple peaks, in \BETSI, it is accompanied by several small modes 
emerging and a small shoulder near the two major peaks. Consequently, obtaining $\delta$-values for the Se-substituted compounds in their insulating state becomes challenging and loses significance. This hints at some remnant signatures of the charge disproportionation, which accompany the charge-ordered insulator phase, but rules out that the phase transition in \STFI\ and \BETSI\ is primarily driven by charge ordering.

As a matter of fact, the occurrence of multiple peaks is expected due to the asymmetry of the inner cyclic rings in \STFI, leading to multiple possible combinations of BEDT-STF stacking arrangements. However, the unusual broadening of the peaks near the phase transition  hints to charge fluctuations in \STFI; as these peaks are resolved again when the crystal is cooled down further. These fluctuations can potentially arise for various reasons: inhomogeneous domains near phase transition due to asymmetric Se arrangement and/or proximity to competing interactions near the phase transition are highly plausible.

\begin{figure*}
	\centering
    	\includegraphics[width=0.85\linewidth]{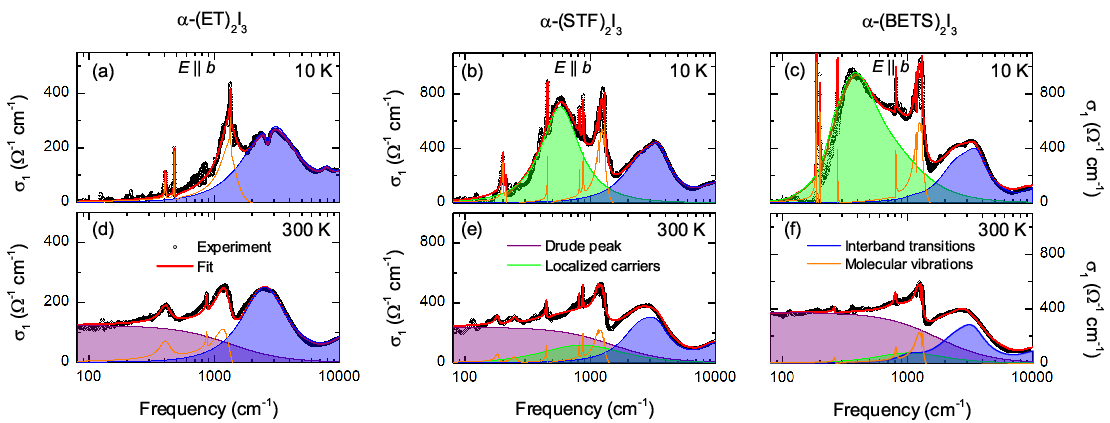}
    	\caption{Optical conductivity fitted by the Drude-Lorentz oscillator model with inclusion of Fano resonance peaks for 10~K (upper row) and 300~K (lower panels) for the polarization parallel to the $b$-axis: (a,d)~\ETI, (b,e)~\STFI{} and (c,f)~\BETSI. The most distinct feature at low temperatures for \STFI\ and \BETSI\ is the strong absorption peak (green) at low energies.}
    \label{fig:Fig5}
\end{figure*}

\subsection{Broadband infrared spectroscopy}

Broadband infrared spectroscopy is the superior method to probe electronic properties of BEDT-TTF-based charge-transfer salts, particularly across phase transitions, and to detect changes in their underlying electronic states with temperature, physical, and chemical pressure \cite{ivek2011electrodynamic, uykur2019optical, dressel2004optical}.
The in-plane optical conductivity of \ETI, \STFI, and \BETSI\ is presented in Fig.~\ref{fig:Fig4}. Previous optical studies were restricted to relatively high energies (above 60~meV or 500~\cm) and our current report is consistent with these reports in this energy range \cite{inokuchi1993electrical, inokuchi1995electrical, naito2020band}.
The optical conductivity combines electronic responses, i.e., inter- and intraband transitions,  and molecular excitations, i.e., molecular vibrations from the bonds of the organic molecules, which may be coupled to the electronic background via emv-coupling, leading to asymmetric line shapes. With \ETI{} being one of the most extensively studied compounds in the family of organic conductors, its optical response is well understood: (1)~when \ETI\ is metallic, the low-energy response (80-1500 \cm) is determined by an overdamped Drude peak, which strongly couples to the $\nu_3$ vibrational mode, centered around 1300 \cm; (2)~the in-plane anisotropy is not very large; (3)~below the metal-insulator phase transition at $T_{\rm MIT}=135$~K, the low-energy contributions from the itinerant carriers are lost abruptly, and spectral weight shifts to the mid-infrared band. Consequently, the multiple molecular vibrations in that region, previously screened by the Drude response, now become sharp and prominent \cite{ivek2011electrodynamic, uykur2019optical}.

Based on our understanding of the electrodynamic properties of \ETI, we now aim to investigate how Se substitution affects the optical response of \STFI\ and \BETSI. By comparing the absolute values of the optical reflectivity (Fig.~S3) and conductivity (Fig.~\ref{fig:Fig4}) of the three compounds well in the metallic state (150 to 300~K), it is evident that the systems become more metallic when moving from ET to BETS by Se substitution. At lower temperatures, close to \TMIT, a strong absorption peak appears in the far-infrared spectra of the substituted compounds, centered around 600~\cm\ for \STFI{} and 450~\cm\ for \BETSI\ (indicated by the black arrows in Fig.~\ref{fig:Fig4}). This feature, completely absent in \ETI, becomes even more prominent below \TMIT.
Attempts to describe $\sigma_1(\omega)$  as presented in Figs.~\ref{fig:Fig5} and S4, (the low-energy response is presented in Figs.~S8 and S9) by Drude-Lorentz fits reveal that this absorption peak at finite frequencies is present already well above the metal-insulator transition (even at $T=300$~K). The presence of this absorption feature and complete localization in the insulating state of \STFI\ and \BETSI\ evidences drastic changes in the charge dynamics of the itinerant carriers that are involved in the phase transition, hinting at a distinct driving force of the MIT in these two compounds, clearly different from the one in \ETI. At $T=10$~K, we obtain the energy gap ($\Delta E_{\rm gap}$) in the insulating state of approximately 600~\cm{} (75 meV), 271~\cm{} (33.5 meV) and 183~\cm{} (22.6 meV) for \ETI, \STFI{} and \BETSI, respectively. The values do not differ for the two polarization axes of the conductivity plane and are consistent with previous reports for \ETI{} \cite{ivek2011electrodynamic}. The gap is significantly reduced by partial Se substitution towards \STFI, while the complete substitution of the inner S atoms in \BETSI\ only leads to a comparatively small decrease in the energy gap.

\begin{figure}
	\centering
    	\includegraphics[width=1\linewidth]{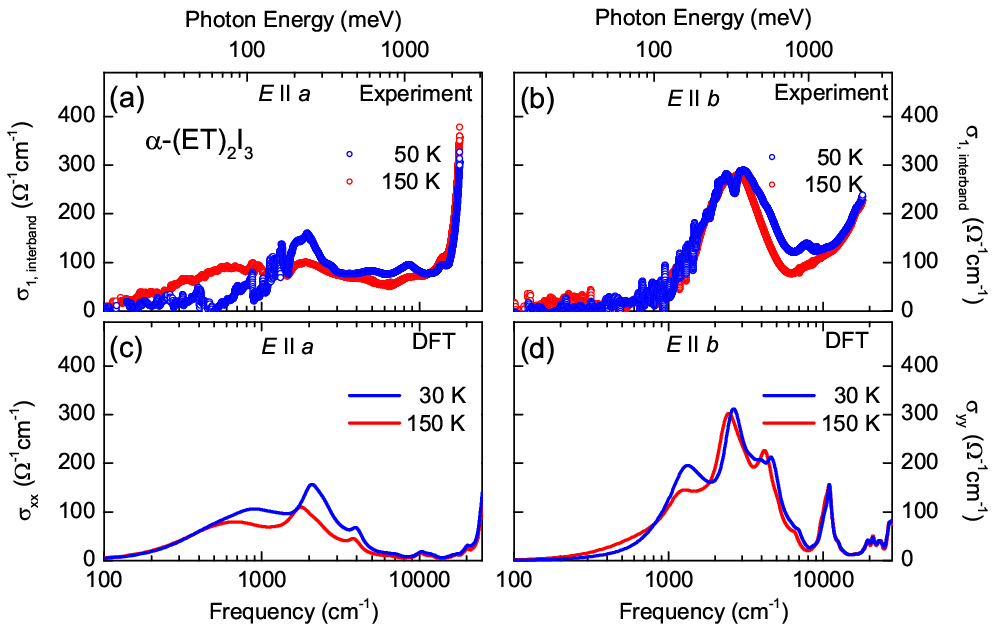}
    	\caption{Comparison of the optical conductivity of \ETI\ measured along the $a$ and $b$ directions with the calculated optical response: (a,b)~The interband transitions of metallic and insulating states are obtained by subtracting the Drude contribution and molecular vibrations from the total optical conductivity. (c,d)~The frequency-dependent conductivity is calculated by density-functional-theory (DFT). The interband transitions are well reproduced except for some deviations along $E \parallel a$ at low temperature, related to the charge ordering.}
    \label{fig:Fig6}
\end{figure}

\begin{figure*}
	\centering
    	\includegraphics[width=0.65\textwidth]{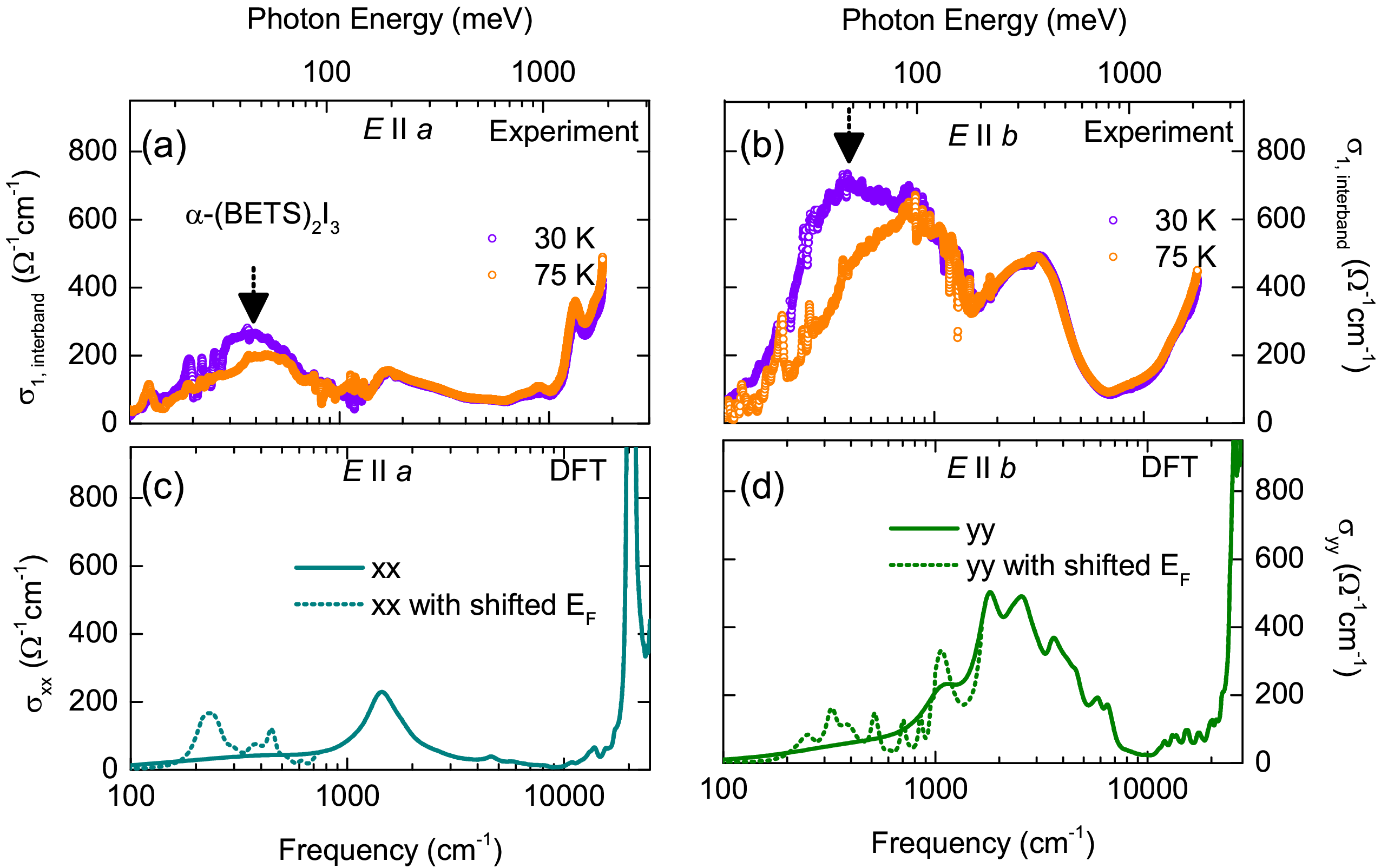}
    	\caption{Comparison of the theoretical and experimental optical conductivity of \BETSI: (a,b)~The interband transitions of the metallic and insulating states of \BETSI\ are obtained by subtracting the Drude-type contribution of the itinerant carriers and the Lorentz-type molecular vibrations from the measured spectra. (c,d)~Interband optical transitions of \BETSI{} calculated by DFT. No major peaks are present below 1000~\cm. The dashed lines represent the spectra obtained by shifting the Fermi level into the energy gap to simulate an insulating state and examine the changes in interband transitions. Still, the experimentally observed low-energy absorption bands, marked by black arrows, are not reproduced.}
    \label{fig:Fig7}
\end{figure*}

The low-energy molecular vibrations in the spectral range between 100 and 1000~\cm, which appear in \ETI\ as charge ordering sets in, become prominent here as well as below the MIT, since they are no longer screened by the free conduction electrons evidenced by the vanishing of the far-infrared Drude peak.
In addition to the absorption peak in the far-infrared region, Se substitution results in a growing anisotropy in the optical response within the conduction plane, which is not limited to the region of the phase transition but observed at all temperatures.

\subsection{Interband transitions}

\begin{table*}
\centering
\caption{Energy gaps ($\Delta E_{\rm gap}$) estimated from the linear extrapolation of the optical conductivity of: \ETI, \STFI, and \BETSI. The ratio of the plasma frequency obtained from the experimental spectra and from DFT calculations is a measure of the effective correlation strength.}
\begin{tabular}{c@{\hspace{0.5cm}}c@{\hspace{1cm}}c@{\hspace{1cm}}c@{\hspace{1cm}}c@{\hspace{0.5cm}}c}
\toprule
Sample  &	$\Delta E_{\rm gap}$ (meV)	&   &  $\omega_{p, \rm exp.}$~(meV)  & $\omega_{p, \rm theo.}$~(meV) & $\omega_{p, \rm exp.}^2$/$\omega_{p, \rm theo.}^2$\\
\midrule
\ETI{} 		&	 75		&	$E \parallel b$ 	&	237.59 (150 K)  &	252	&	0.88 \\
 	   		&			&	$E \parallel a$ 	&	105.61 (150 K)  &	184 &	0.32	\\		
\STFI{} 	& 	33.5	& 	$E \parallel b$  	&	353.84 (75 K) 	& 	 		\\
	  		& 			&	$E \parallel a$ 	&	184.49 (75 K)   & 	 		\\
\BETSI{} 	&	22.6	&	$E \parallel b$ 	& 	391.85 (75 K)	&	745	& 	0.27	\\
    		&			& 	$E \parallel a$ 	& 	199.75 (75 K)	&	303 &	0.43	\\
\bottomrule
\end{tabular}
\label{tab:table1}
\end{table*}

Band structure calculations help bridging theory and experiment when analyzing the optical response; based on the optimized structure, the density of states and eventually the frequency-dependent conductivity is evaluated. The agreement or disagreement between the experiment and theory provides insights into electronic correlations and their effect on the electronic band structure. The experimental interband transitions are obtained by subtracting the Drude contribution and molecular vibrations extracted from the fits of $\sigma_1(\omega)$. The theoretical conductivities of \ETI{} and \BETSI{} are calculated from the band structures reported in Ref.~\onlinecite{kitou2021ambient}. It is important to highlight that most of the theoretical studies are limited to the symmetric \ETI{} and \BETSI, as the asymmetric BEDT-STF molecules possess additive complications due to multiple combinations in the stacking arrangement within a unit cell; for that reason attempts have been made of averaging the orientation combinations and treating the asymmetry uniformly across the crystal system \cite{naito2020exotic, naito2022theoretical}. For simplified discussion, we only consider the symmetric \ETI\ and \BETSI\ systems in our study.

Fig.~\ref{fig:Fig6} demonstrates the very good agreement between the DFT-based and experimentally obtained interband optical transitions of \ETI. Note that the frequency axis of the theoretical conductivities is multiplied by a factor of 1.6 to match the energy of the low-energy peaks with the experimentally observed ones. Such a rescaling indicates the renormalization of the low-energy electronic band structure possibly as an effect of correlations. Since charge ordering is not completely captured by the DFT calculations, the low-temperature $E \parallel a$ response is not well described and can be used to estimate the gap that opens by the charge ordering from the experimental response.

For \BETSI, the interband transitions above 1000~\cm\ are well reproduced by DFT, while the significant spectral weight in the low-energy band cannot be explained by interband transitions, as there are no major changes in the band structure, compared to \ETI\ (see Fig.~\ref{fig:Fig6}). Since the energy gap lies very close to $E_F$, even slight variations in the sample's Fermi level relative to the DFT-calculated value can significantly alter the ground state. To account for this, we shifted $E_F$ downward by 20 meV to simulate a band-insulating phase. However, this barely affects the interband transitions, as illustrated in Fig.~\ref{fig:Fig7}(c,d), and, therefore, fails to reproduce the low-energy absorption. Furthermore, the band structure of \BETSI\ does not change between 30 and 80~K; and we do not observe abrupt changes in the experimental interband transitions in accord with Refs.~\onlinecite{ohki2023gap, kitou2021ambient}. Any effects of strong correlations similar to \ETI{} can be disregarded, as we can neither identify strong renormalization nor changes in interband transitions which typically occur when correlation-driven effects cause an insulator phase. This identifies the low-energy response as part of intraband transitions, implying that the response of the itinerant carriers in \BETSI{} and \STFI{} is not a conventional Drude-behavior but resembles some dynamic localization in the metallic states, possibly led by strong electron-phonon coupling and influence of Bosonic modes on the itinerant carriers \cite{keski2024quantum, rammal2024transient}. This interpretation is further supported by the drastic spectral weight transfer between the Drude peak (free carriers) and localized charge carriers demonstrated in Fig. S9. 

The strength of electronic correlations is estimated by comparing the experimental and DFT-based plasma frequencies, $\omega_{p, \rm exp.}^2$/$\omega_{p, \rm theo.}^2$, as summarized in Table ~\ref{tab:table1} \cite{shao2019optical, qazilbash2009electronic}. This ratio is close to 1 for uncorrelated systems, while it approaches 0 in the case of a Mott insulating state. $\omega_{p, \rm exp.}^2$ is calculated from spectral weight integration of the intraband response \citep{dressel2002electrodynamics}. Interestingly, the correlation strength in \ETI\ exhibits pronounced anisotropy, with the \textit{b}-plane being nearly uncorrelated, while along the \textit{a}-axis, i.e., the direction in which charge ordering is observed, correlations increase significantly. In contrast, \BETSI\ displays relatively similar and moderate correlations along both directions.

\section{Discussion}

\begin{figure*}
    \centering
        \includegraphics[width=1\linewidth]{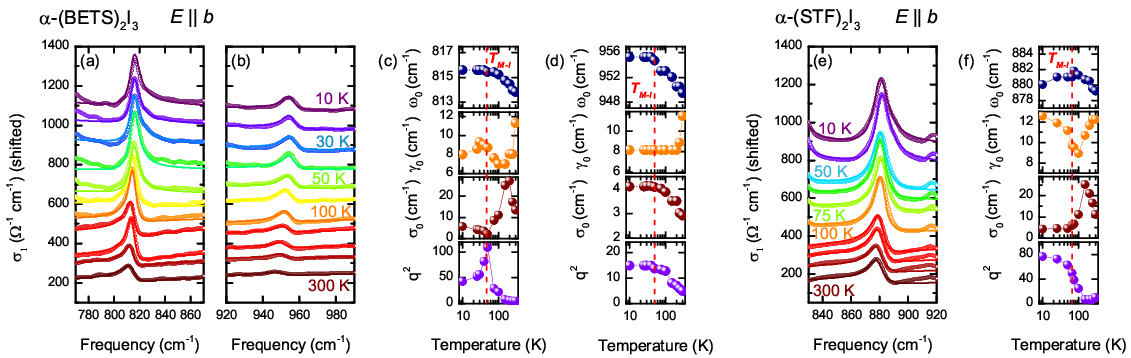}
        \caption{Temperature evolution of the optical signature of selected molecular vibrations. These modes exhibit unusual response with temperature, which can be captured by a Fano fit. The respective Fano parameters are plotted as a function of temperature for (a-d) \BETSI{} and (e-f) \STFI.}
    \label{fig:Fig8}
\end{figure*}

In general, external pressure reduces the effects of correlations by increasing the bandwidth for BEDT-TTF-based organic conductors. In the case of \ETI, hydrostatic pressure causes a linear decrease of $T_{\rm MIT}$, with retention of the optical signatures of a charge-ordered insulator up to high pressure \cite{beyer2016pressure}. Here, no fingerprint of charge localization, as observed in the Se-substituted compounds, can be identified up to 4 GPa \cite{beyer2016pressure, uykur2019optical}.

On the other hand, our analysis of band structures and interband transitions in Section III(D) and Section V of the Supplemental material \cite{SM} reveals that the replacement of S by the larger Se or Te atoms may increase the bandwidth, however, the experimental observation of a prominent absorption feature at low energy, points to a more complex underlying mechanism for the insulating state in \STFI\ and \BETSI. Already in the metallic regime, i.e., for temperatures between 75 to 300 K, the low-energy dynamics of \STFI{} and \BETSI{} can be fitted  best by a combination of an over-damped Drude peak and a Lorentz peak to account for the additional absorption. At low temperatures, the differences become significant: along both polarization directions, a pronounced absorption band appears in the low-energy response of \STFI{} and \BETSI; no comparable excitation can be identified in \ETI.

In accordance with our transport measurements and broadband optical investigations, focusing on the charge-sensitive molecular vibrations and electronic response, we can rule out a thermally activated metallic states (as observed for narrow-gap semiconductors), disorder-driven or correlation-mediated charge-ordered insulating states for \STFI\ and \BETSI. 

The evolution of the low-temperature conductivities of the two Se-containing $\alpha$-salts qualitatively resembles the formation of density-wave states, similar to observations in quasi-one-dimensional compounds such as TTF- or TMTSF-based systems  \cite{GrunerBook,Donovan1994SDW,Gorshunov1994CDW,Schwartz1995CDW,Degiorgi1996SDW,Vescoli1999SDW,zwick2000optical}. In these compounds, an energy gap opens at $E_F$ due to nesting, observed as a pronounced conductivity peak, and is accompanied by structural modulation or charge ordering. First, we recall that \ETI\ does not undergo a nesting-driven charge-density wave transition at $T_{\rm MIT}=135$~K, but pronounced charge ordering on the BEDT-TTF molecules \cite{Dressel1994aET2I3,Tomic2015Ferro}. Moreover, no structural changes which may evidence a density wave formation have been reported, at the phase transition for \STFI{} and \BETSI. In general, robust evidence for density wave formation in two-dimensional organic conductors is missing, and even for the small number of candidates, such as $\alpha$-(BEDT-TTF)$_2M$Hg(SCN)$_4$, experimental indications are sparse \cite{Maki2003CDW,Dressel03b,Drichko06,Drichko06b,Kartsovnik2014CDW}. Although future research might disclose further indications for density wave formation in the quasi-two-dimensional BEDT-TTF-based compounds, currently, we conclude that the observed bands in the low-frequency absorption of \STFI{} and \BETSI{} are not related to a density wave formation. In other words, the finite-energy absorption peak is not due to modifications in the band structure or density of states.

Molecular vibrations in the BEDT-TTF-based compounds are sensitive to changes in the electronic response and act as local probe to investigate minute changes. Some of the molecular vibrations of \STFI{} and \BETSI{} in the range of 800 to 950~\cm{} demonstrate strong asymmetry with a distinct temperature evolution. While such an asymmetric Fano line shape of molecular vibrations in infrared spectra is quite common \cite{iakutkina2021charge, Yakushi2012}, the peculiar temperature evolution of these modes, deviating from the usual phonon hardening, can be viewed as an indication of modifications in the underlying electronic structure. For \BETSI, as demonstrated in Fig.~\ref{fig:Fig8}(a-d), the anomalies in the Fano resonance fit parameters coincide with the metal-insulator phase transition temperature \TMIT, while for \STFI, the changes start slightly above the transition temperature [Fig.~\ref{fig:Fig8}(e-f)], which is consistent with the changes in low-energy electronic response (see Fig.~S9 for the intraband response) and potentially arises from the charge fluctuations in \STFI. This observation suggests that the most probable reason for the phase transition in \STFI\ and \BETSI\ is the direct modification of the itinerant carriers response. 

From our recent ellipsometry study \cite{tiwari2025nature}, we concluded a continuous change in the metallic state of Se-substituted systems arising from a complex interplay of chemical substitution (acting as chemical pressure) and the accompanying orbital alterations. 

\begin{figure}[b]
    \centering
        \includegraphics[width=0.8\linewidth]{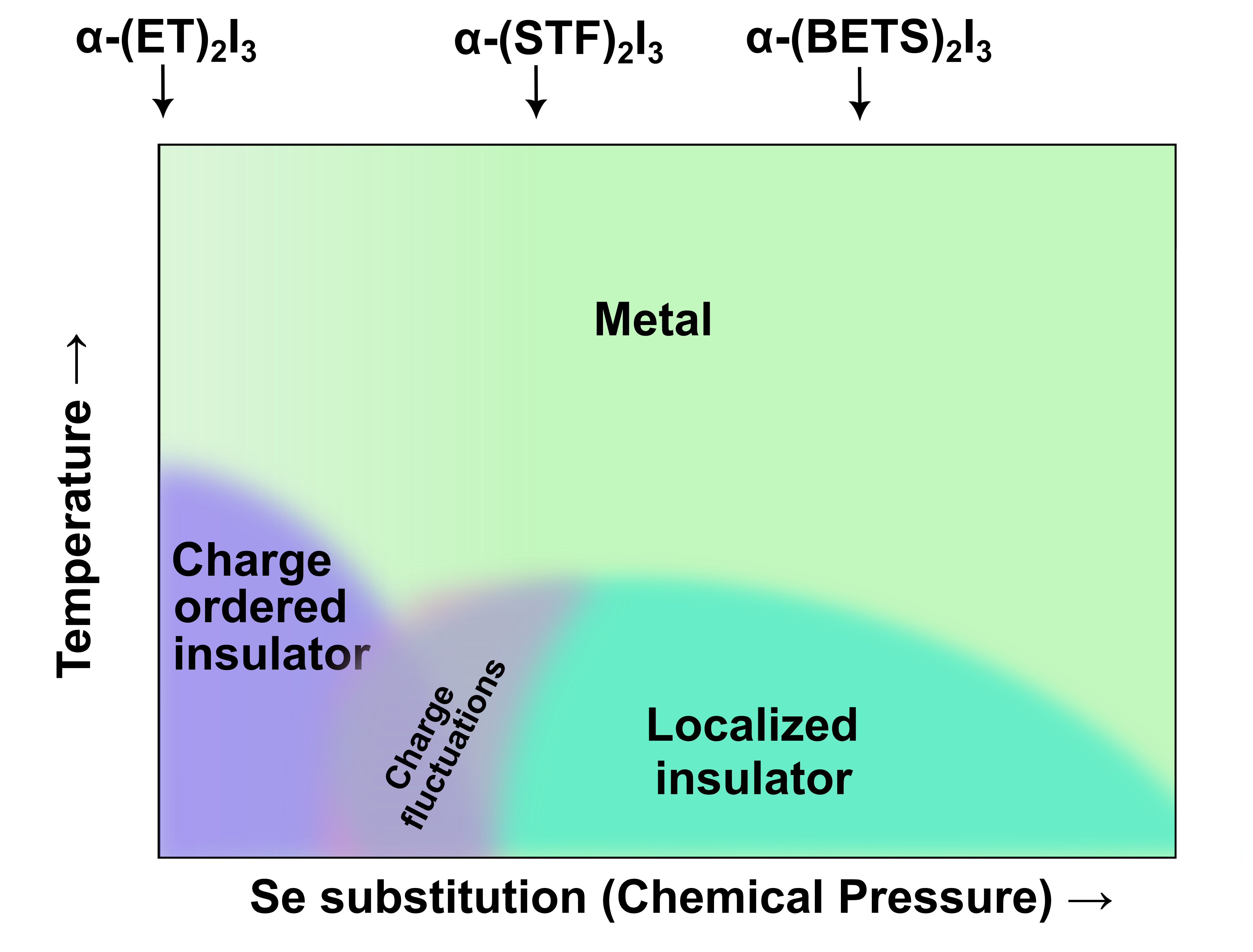}
        \caption{Phase diagram proposed on the basis of our optical results, unveiling the charge localization-induced phase transition observed in \STFI{} and \BETSI. The signatures from both charge-ordered insulator state and localized charge insulator phase for \STFI{} places it in very close to the cross-over regime, which is accompanied by charge fluctuations.}
    \label{fig:Fig9}
\end{figure} 

Taking into account the above results and interpretations, we suggest that upon Se substitution, correlation-driven charge-ordering is suppressed due to more isotropic and moderate correlations (while \ETI\ shows highly anisotropic correlations), and fundamental interactions between electrons and bosonic modes become more important. Since there are no indications of magnetic moments, magnetic order, fluctuation or spin-orbit coupling in these solids, the decisive mechanism for the localization of charge carriers is attributed to the lattice-related interactions, i.e., the electron-phonon coupling. A similar modification of the itinerant carriers' response has been previously observed as dynamic localization in organic semiconductors as well \cite{fratini2016transient, rammal2024transient}. 

Since the increased chemical pressure (Se substitution), isotropic correlations, and increased effects of electron-phonon coupling in \STFI\ and \BETSI\ occur simultaneously, their influence or inter-dependence cannot be fully disentangled. Based on our optical response, we propose the phase diagram displayed in Fig.~\ref{fig:Fig9}, demonstrating the transformation from a charge-ordered insulating state in \ETI\ to a localized insulating one in Se-substituted counterparts, due to effects of strong electron-phonon coupling. Additionally, it has to be noted, that the charge fluctuations near the phase transition observed in \STFI{} cannot be attributed solely to asymmetry in the molecular structure, as it exhibits remnant signatures of both charge ordering and strong low-energy localization. This indicates that charge fluctuations arise due to the proximity to crossover of the two different phase transition mechanisms.

Additional strain measurements on these $\alpha$-phase compounds could enable tuning between the charge-ordered and localized insulating states, influenced by anisotropic correlations. This also motivates further synthesis and investigation of less-studied Te-based organic charge-transfer salts, aiming to explore the complete range of chemical pressure effects.  

\section{Conclusion}

In our comparative optical study of the three $\alpha$-phase organic conductors, we investigate the effects of chemical substitution on the electrodynamic properties. We unveil a low-energy charge localization, responsible for metal-to-insulator phase transition in the Se-substituted compounds. The presence of larger Se atoms exerts a form of chemical pressure, however, the resulting effects differ significantly from the effects of hydrostatic pressure applied on \ETI. The charge disproportionation in the Se-substituted samples does not exhibit signs of charge ordering at the phase transition. Altogether, this points towards a shift from a correlation-driven charge-ordered insulating electronic ground state observed in \ETI{} to a more complex interactive state, dominated by the itinerant carriers coupled to phonons, resulting in localized carriers and an eventual insulating state in \STFI{} and \BETSI.

\section*{Acknowledgement}

We thank Andrej Pustogow for fruitful discussions. We thank Gabriele Untereiner for the technical support.  The work at University of Stuttgart is supported by the Deutsche Forschungsgemeinschaft (DFG) DR 228/39-3 and DR 228/68-1. 

\bibliography{ref}

\end{document}


\title{{\Large Supplemental Material }\\
\vspace*{1mm}
Charge-localization-driven metal-insulator phase transition \\
in layered molecular conductors}

\author{Savita Priya}
\affiliation{1. Physikalisches Institut, Universit\"at Stuttgart, Pfaffenwaldring 57,
70569 Stuttgart, Germany}
\author{Maxim Wenzel}
\affiliation{1. Physikalisches Institut, Universit\"at Stuttgart, Pfaffenwaldring 57,
70569 Stuttgart, Germany}
\author{Olga Iakutkina}
\affiliation{1. Physikalisches Institut, Universit\"at Stuttgart, Pfaffenwaldring 57,
70569 Stuttgart, Germany}
\author{Marvin Schmidt}
\affiliation{1. Physikalisches Institut, Universit\"at Stuttgart, Pfaffenwaldring 57,
70569 Stuttgart, Germany}
\author{Christian Prange}
\affiliation{1. Physikalisches Institut, Universit\"at Stuttgart, Pfaffenwaldring 57,
70569 Stuttgart, Germany}
\author{\\Dieter Schweitzer}
\affiliation{1. Physikalisches Institut, Universit\"at Stuttgart, Pfaffenwaldring 57,
70569 Stuttgart, Germany}
\author{Yohei Saito}
\affiliation{1. Physikalisches Institut, Universit\"at Stuttgart, Pfaffenwaldring 57,
70569 Stuttgart, Germany}
\author{Reizo Kato}
\affiliation{RIKEN, 2-1, Hirosawa, Wako-shi, Saitama 351-0198, Japan}
\author{Koichi Hiraki}
\affiliation{Center for Integrated Science and Humanities, Fukushima Medical University, Fukushima 960-1295, Japan}
\author{Martin Dressel}
\affiliation{1. Physikalisches Institut, Universit\"at Stuttgart, Pfaffenwaldring 57,
70569 Stuttgart, Germany}

\date{\today}
\maketitle

\section{Transport measurements}

\subsection{Thermal activation}

\begin{figure}
    \centering
      \includegraphics[width=1\columnwidth]{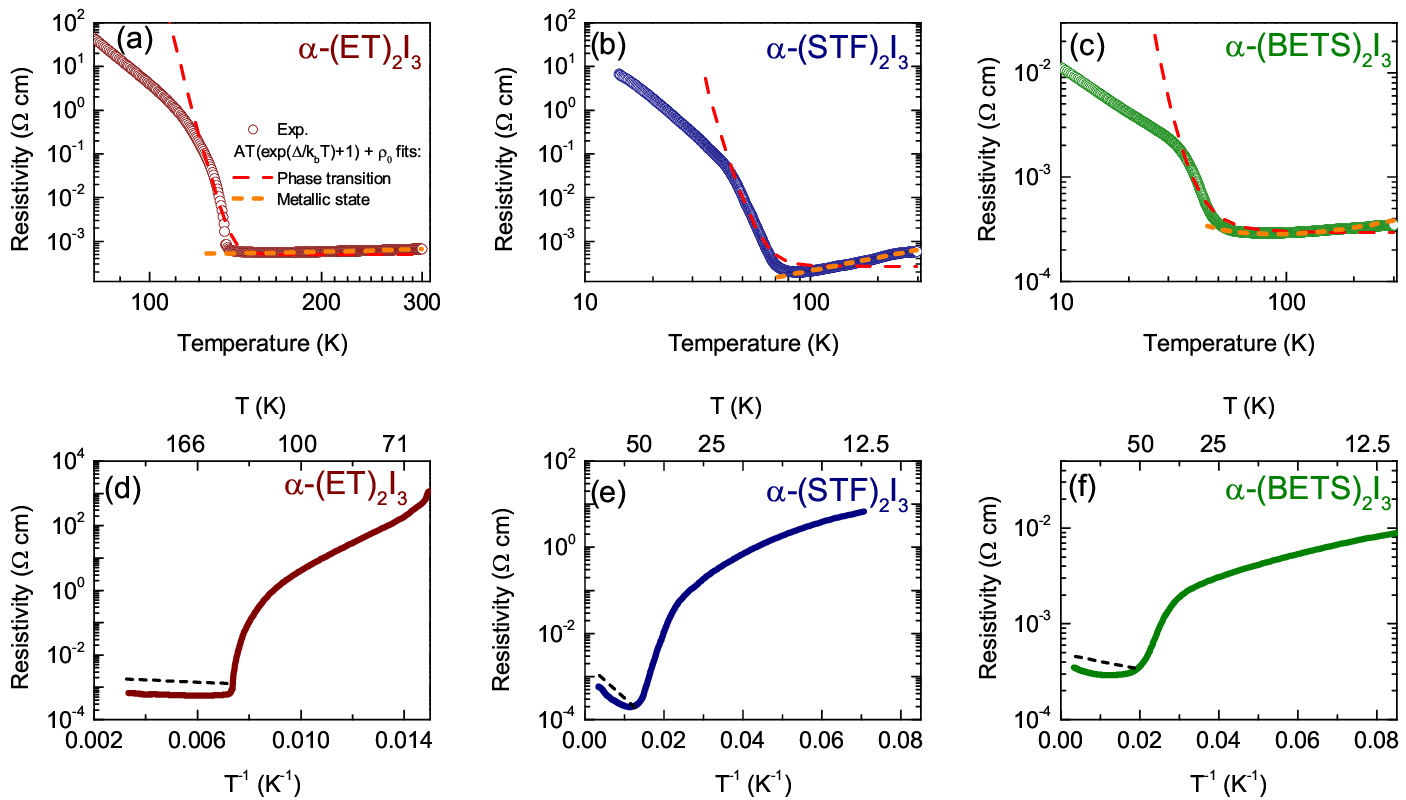}
        \caption{Electrical resistivity of $\alpha$-(ET)$_2$I$_3$, $\alpha$-(STF)$_2$I$_3$, and $\alpha$-(BETS)$_2$I$_3$ as a function of temperature. (a-c) Double-logarithmic representation and fits by Arrhenius equation, (d-f) Arrhenius presentation (the dashed straight lines are guides to the eye).}
    \label{fig:FigS1}
\end{figure}

In Fig.~\ref{fig:FigS1}(a-c) we plot the temperature dependence of the dc resistivity of \ETI, \STFI, and \BETSI\ on a double-logarithmic scale, where ET=BEDT-TTF stands for bis\-(ethylene\-di\-thio)tetra\-thia\-fulvalene, STF=BEDT-STF denotes bis\-(ethylene\-di\-thio)selena\-thia\-fulvalene,  and BETS=BEDT-TSF stands for bis\-(ethylene\-di\-thio)tetra\-selena\-fulvalene. At high temperatures, a metallic regime is identified in all three cases, where $\rho(T)$ changes only marginally upon cooling down to the metal-insulator phase transition; for \STFI, the metallic temperature dependence is most pronounced.

Thermally activated conduction, with an activation energy ($\Delta$), considering electron-phonon coupling and Fermi-Dirac statistics (modified Arrhenius behavior) is given as following: 
\begin{equation}
\rho(T) = A T \left\{ \exp \left( \frac{\Delta}{k_BT} + 1 \right) \right\} + \rho_0 \quad , 
\label{eq:S1}
\end{equation}
where, $\rho_0$ is a prefactor, and $C$ indicates an offset \cite{baranovski2006charge}. While this equation satisfactorily reproduces the resistivity in the metallic state (orange dashed line in Figs.~\ref{fig:FigS1}(a-c)), it cannot reproduce the sharp change at the phase transition for any of the three $\alpha$-phase compounds (red dashed line in Figs.~\ref{fig:FigS1}(a-c)).

In order to gain insight into the physical mechanism of the charge transport at the phase transition, Figs.~\ref{fig:FigS1}(d-f) display the same data in the conventional Arrhenius form, i.e. $\log\{\rho(T)\}$ {\it vs.} $T^{-1}$, which allows to readily identify a thermally activated behavior:
\begin{equation}
\rho(T) \propto  \exp \left\{ \frac{\Delta}{k_BT} \right\}  \quad .
\label{eq:S2}
\end{equation}
Although, we may identify a straight line in this presentation for narrow temperature ranges, as indicated by the dashed lines, we cannot conclude a simple activated behavior with a fixed gap $\Delta$ in any of the three cases. The slope decreases at low temperatures in metallic state. Hence, we can rule out a thermally activated transport  in \STFI, and \BETSI.

\newpage

\newpage

\subsection{Variable-range-hopping (VRH) mechanism}
Since the resistivity does not follow a thermally activated  behavior, the charge transport could take place via variable-range-hopping (VRH). This model typically describes effects of incoherent transport due to intrinsic disorder or the effects of introduced defects \cite{mott2012electronic}. Depending on the dimensionality, $d$ of the system, VRH model is given as:

\begin{equation}
\rho(T) \propto \exp\left\{ \left( \frac{T_0}{T} \right)^{\frac{1}{d+1}}  \right\} \quad .
\label{eq:S3}
\end{equation}

\begin{figure}[h]
    \centering
        \includegraphics[width=0.8\linewidth]{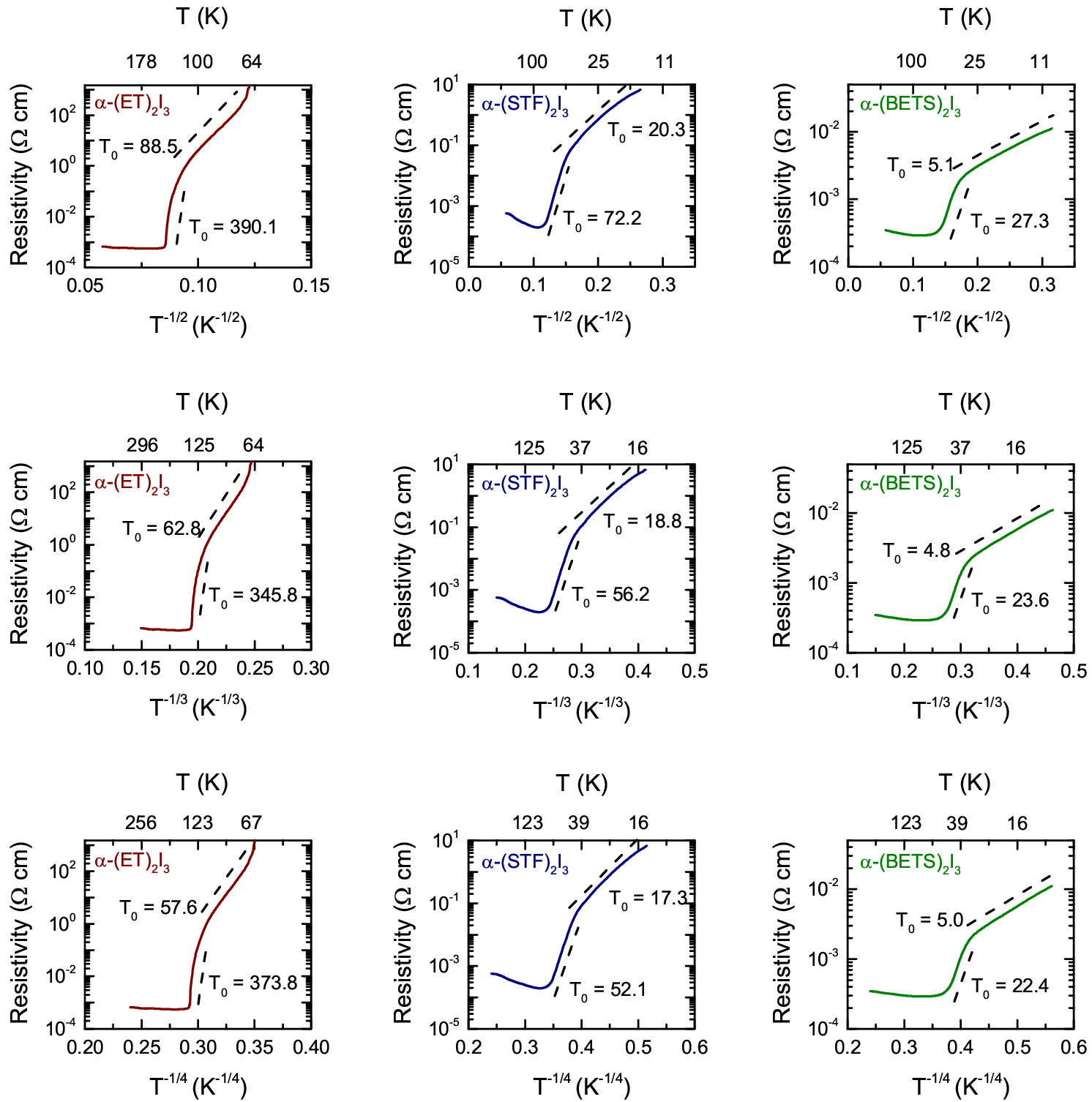}
        \caption{Temperature-dependent resistivity $\rho(T)$ of \ETI{} (left), \STFI{} (center), and \BETSI{} (right) plotted as a function of $T^{-1/2}$ (upper frames), $T^{-1/3}$ (central panels), and $T^{-1/4}$ (lower panels). The dashed lines are guides to the eye, indicating certain regions where the data may be described by the particular lower law.}
    \label{fig:FigS2}
\end{figure}

Although \ETI{} is quasi-two-dimensional, Se-substitution can lead to changes in the dimensionality. Hence, we plot log$\{\rho\}$ $vs.$ $T^{-1/2}$, $T^{-1/3}$, and $T^{-1/4}$ for $d =1$, 2, 3 in Fig.~\ref{fig:FigS2}.
Below the metal-insulator phase transitions at 135, 66, and 50~K, for for \ETI, \STFI, and \BETSI, respectively, we can identify another change in slope at 120, 50, and 30~K, respectively. This rather intriguing behavior does not support the inherent disorder-driven phase transition scenario.

\clearpage

\section{Reflectivity: in-plane}
\begin{figure*}[h]
    \centering
        \includegraphics[width=1\linewidth]{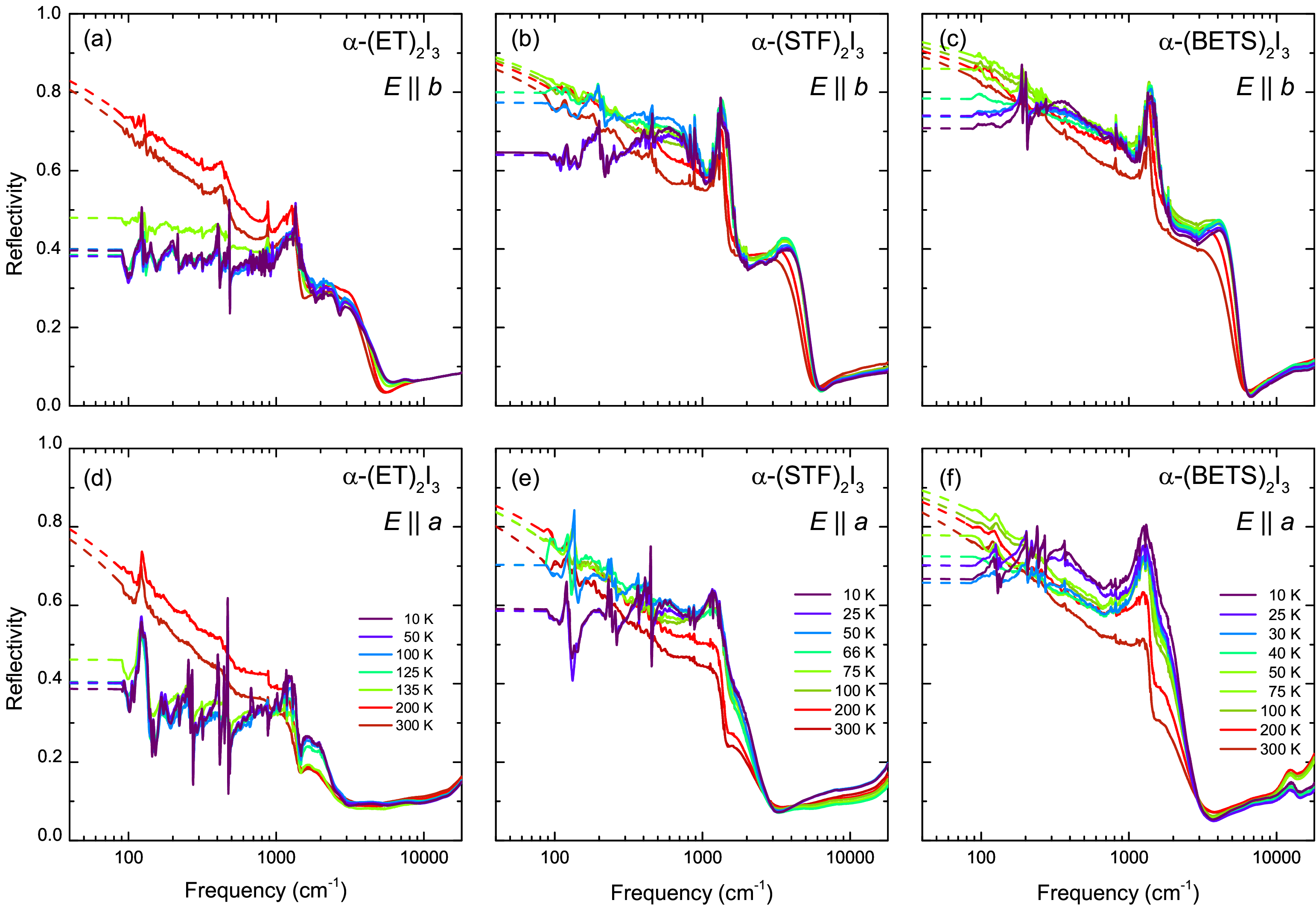}
        \caption{Optical reflectivity recorded at different temperatures between 10 and 300~K for two different in-plane polarization axes ($E \parallel a$ and $E \parallel b$) for (a,d)~\ETI, (b,e)~\STFI, and (c,f)~\BETSI. Note the logarithmic frequency axes. The low-energy extrapolation (below 80 \cm) is indicated by dashed lines.}
    \label{fig:FigS3}
\end{figure*}

In Fig.~\ref{fig:FigS3} the temperature evolution of the in-plane optical reflectivity of \ETI, \STFI, and \BETSI\ is plotted for the two polarization directions, $E\parallel b$ (upper panels) and $E\parallel a$ (lower panels). The metal-insulator phase transitions at 135, 66 and 50~K, respectively, can be clearly seen in the drop of the low-frequency reflectivity.

\newpage

\section{Determination of the optical conductivity and decomposition of different contributions}

The experimentally measured reflectivity, plotted in Fig.~\ref{fig:FigS3}, is utilized to calculate the optical conductivity, following Kramers-Kronig transformation and calculation of complex dielectric response. The results for \ETI, \STFI, and \BETSI\ obtained at different temperatures are plotted in Fig.~4 of the main text. For more details see Ref.~\onlinecite{dressel2002electrodynamics}.

The optical conductivity can be resolved in

\begin{itemize}
\item low-energy intra-band transitions (due to the itinerant carriers' response) fitted by a combination of a Drude peak and a Lorentz oscillator,
\item inter-band transitions (finite-frequency absorption bands) fitted by Lorentz oscillators, and
\item sharp asymmetric molecular vibrations of BEDT-TTF, which overlap with the electronic background (Fano resonance peaks) \cite{damascelli1997infrared}.
\end{itemize}
The overall optical conductivity can be fitted by a combination of a Drude term, multiple Lorentz oscillators and Fano resonance peaks to disentangle different contributions systematically.

\begin{equation}
\hat{\sigma}(\omega) =
\underbrace{\hat{\sigma}_{\text{intraband}}}_{\text{Drude + Lorentz}}
+
\underbrace{\hat{\sigma}_{\text{interband}}}_{\text{Lorentz}}
+
\underbrace{\hat{\sigma}_{\text{molecular vibrations}}}_{\text{Fano}}
\label{eq:S3}
\end{equation}

The final fit function is given by:
\begin{equation}
\hat{\sigma}(\omega) = -{\rm i}\omega \varepsilon_0
\left[
-\frac{\omega_p^2}{\omega^2 + {\rm i}\omega/\tau_{\rm Drude}}
+ \frac{\Omega_{({\rm localization})}^2}{\omega_{0}^2 - \omega^2 - {\rm i}\omega\gamma}
\right]
-{\rm i}\omega \varepsilon_0 \sum_k \frac{\Omega_{k({\rm inter})}^2}{\omega_{0,k}^2 - \omega^2 - {\rm i}\omega\gamma_k}
+ {\rm i} \left[\sum_l \sigma_0 \frac{(q -{\rm i})^2}{{\rm i} + \frac{\omega_{l}^2-{\omega_{0,l}^2}}{\gamma_l\omega_{l}}}\right] ,
\label{eq:S4}
\end{equation}
where $\tilde{\sigma}$ is the complex optical conductivity, $\omega_{p}$ indicated the plasma frequency, $\tau_{\rm Drude}$ is the scattering rate of itinerant charge carriers; $\omega_{0}$, $\Omega$ and $\gamma$ are resonance frequency, oscillator strength and width of the Lorentz peak for fitting the localized carriers' response; $\omega_{0,k}$, $\Omega_{k}$ and $\gamma_{k}$ are resonance frequency, oscillator strength and width of the peak for the $k^{th}$ Lorentz peak for the interband contributions; $q$ is the Fano parameter, $\sigma_0$ is the amplitude and $\omega_{l}$ is the resonance frequency of the Fano peaks \cite{damascelli1997infrared}.

\newpage
\begin{figure}
    \centering
        \includegraphics[width=0.6\linewidth]{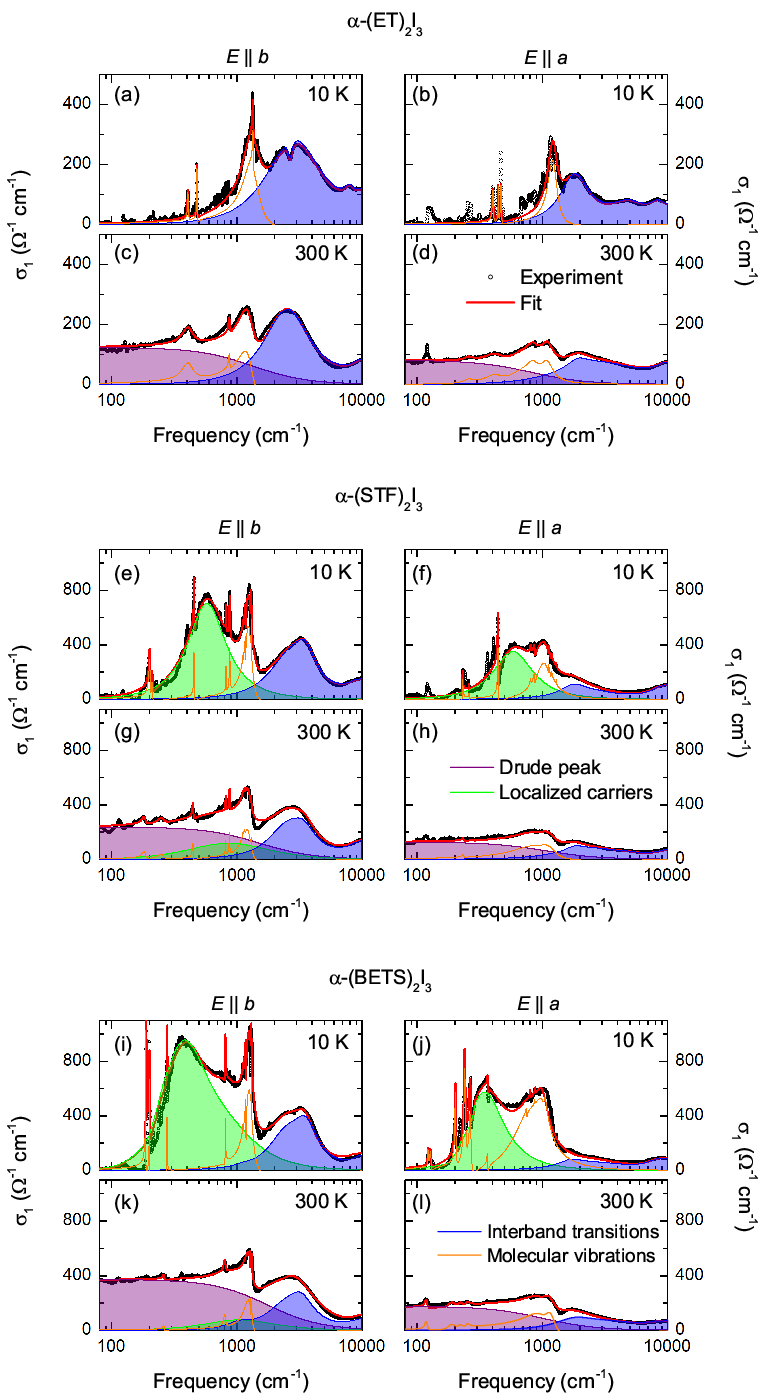}
        \caption{Optical conductivity, $\sigma_1(\omega)$ fitted by the Drude-Lorentz oscillator model with inclusion of Fano resonance peaks for two temperatures $T=10$ and 300 K: (a-d)~\ETI, (e-h)~\STFI{} and (i-f)~\BETSI{} in the two in-plane polarization axes; various peaks decomposed from the fits are represented by different colors. The most distinguishing feature at low temperatures for \STFI{} and \BETSI{} is the strong absorption peak due to localized carriers at low energies.}
    \label{fig:FigS4}
\end{figure}

\newpage

\section{Out-of-plane reflectivity and optical conductivity}
\begin{figure}[h]
    \centering
        \includegraphics[width=0.9\linewidth]{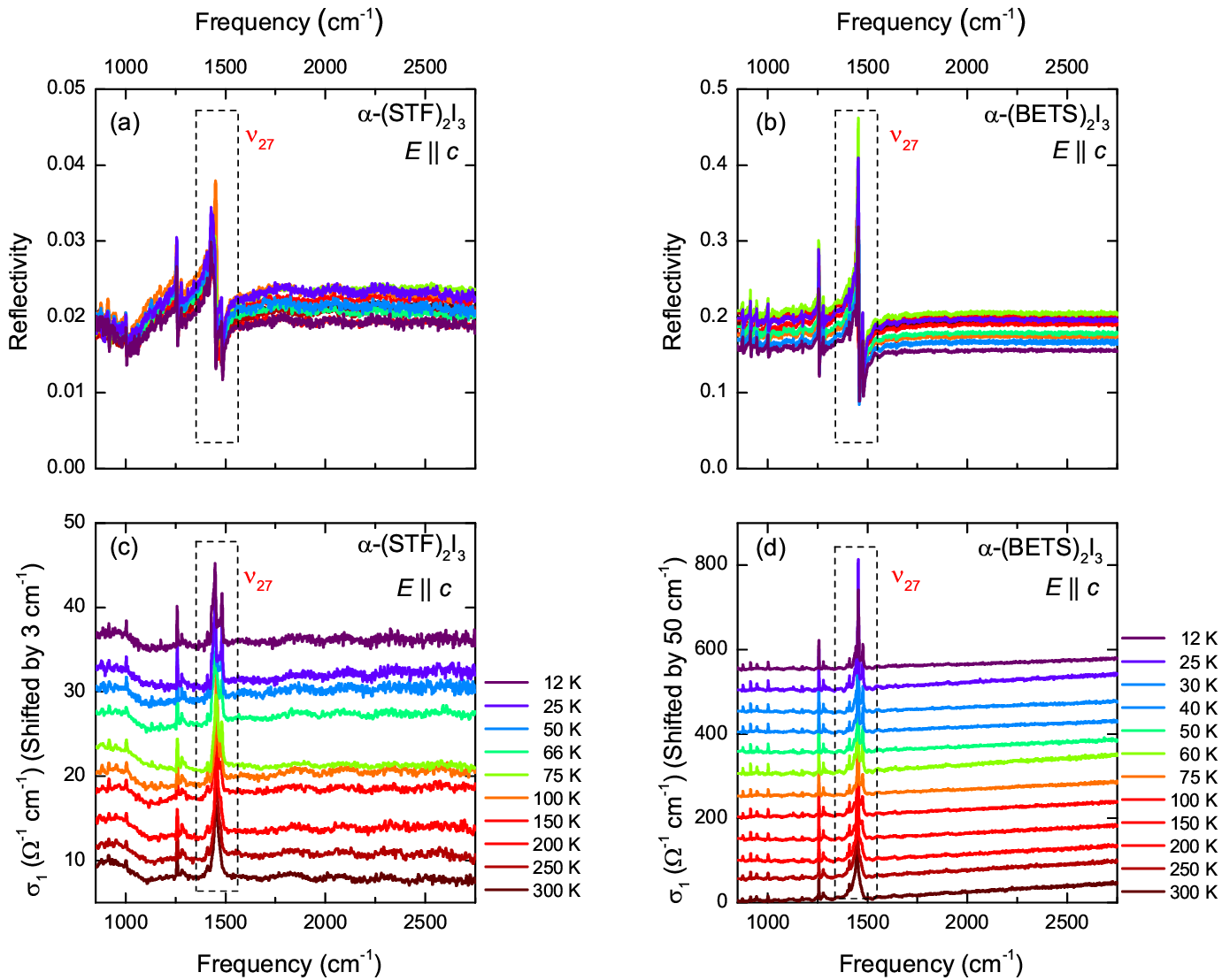}
        \caption{Temperature-dependent out-of-plane (a,b) reflectivity and (c,d) optical conductivity in the region of charge-sensitive molecular vibration, $\nu_{27}$ for \STFI{} and \BETSI. The low absolute values for \STFI{} are due to relatively small area of flat surface along the thickness ($<$ 20 $\mu$m) of the crystal where the measurement was performed.}
    \label{fig:FigS5}
\end{figure}

The out-of-plane response is probed by optical reflection measurements with the electric field polarized perpendicular to the planes ($E\parallel c$), i.e., on the thin sides of the crystals. In Fig.~\ref{fig:FigS5}, the optical reflectivity and conductivity is plotted for different temperatures. Due to experimental challenges, the absolute values for \STFI\ may be incorrect, but this does not affect the frequency and temperature behavior of the vibrational features.

\newpage

\section{Band structures by DFT calculations}

\begin{figure}[h]
    \centering
        \includegraphics[width=1\linewidth]{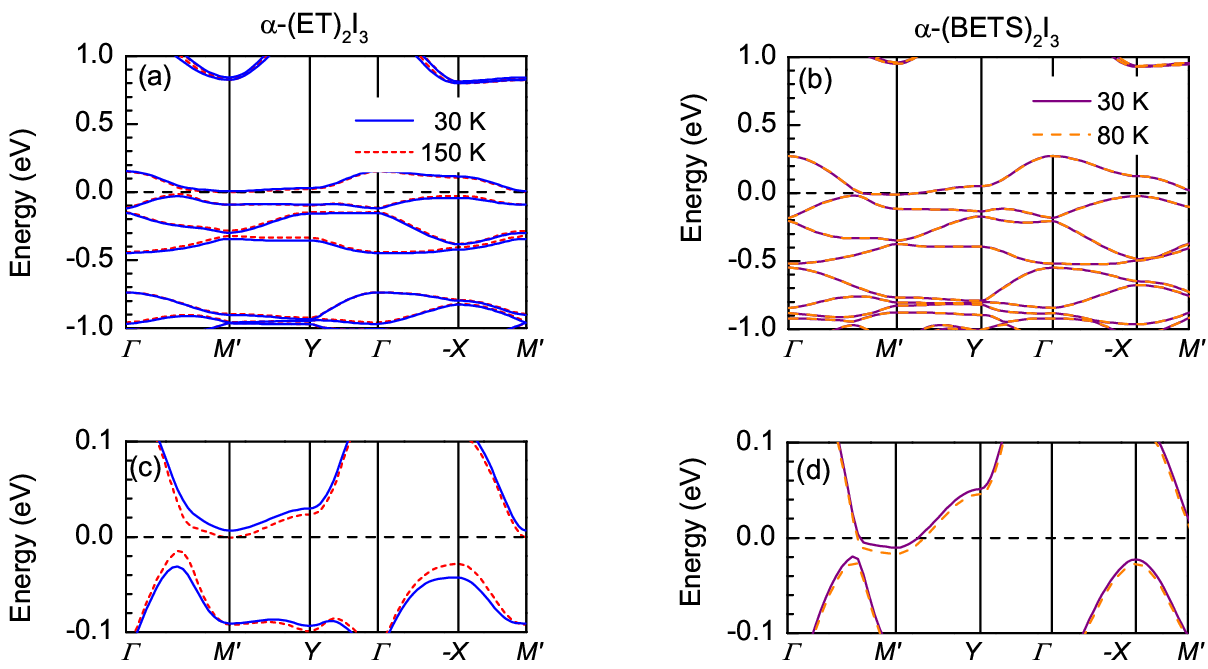}
        \caption{Band structure for (a) \ETI, and (b) \BETSI. The metallic states are indicated by the dashed lines, the insulating states by the solid lines. (c,d) Magnified bandstructure near $E_F$ for (c) \ETI, and (d) \BETSI.}
    \label{fig:FigS6}
\end{figure}

We calculate the band structure of \ETI{} and \BETSI{} using the experimental crystal structures at different temperatures, following Ref. \cite{ohki2023gap}, by Quantum Espresso \cite{giannozzi2009quantum, giannozzi2017advanced}. In the case of \ETI, the phase transition is accompanied by a loss of inversion symmetry, leading to a gap opening in the low-temperature insulating state [Fig.~\ref{fig:FigS6}(a,c)], while the high-temperature band structure shows metallic behavior. For \BETSI, there are no reports of a pronounced change in the  crystal structure at the phase transition. Hence, the band structure is almost identical for the metallic ($T=80$~K) and the insulating states (30 K), with a tiny gap present. The Fermi level lies very close to the gap between the two bands and marginally crosses the top band, making the electronic ground state of \BETSI{} metallic, consistent with the experimental results at temperatures above MIT. The metal-insulator transition at 50~K cannot be readily explained by changes in the bandstructure.

\newpage

\begin{figure}
    \centering
        \includegraphics[width=1\linewidth]{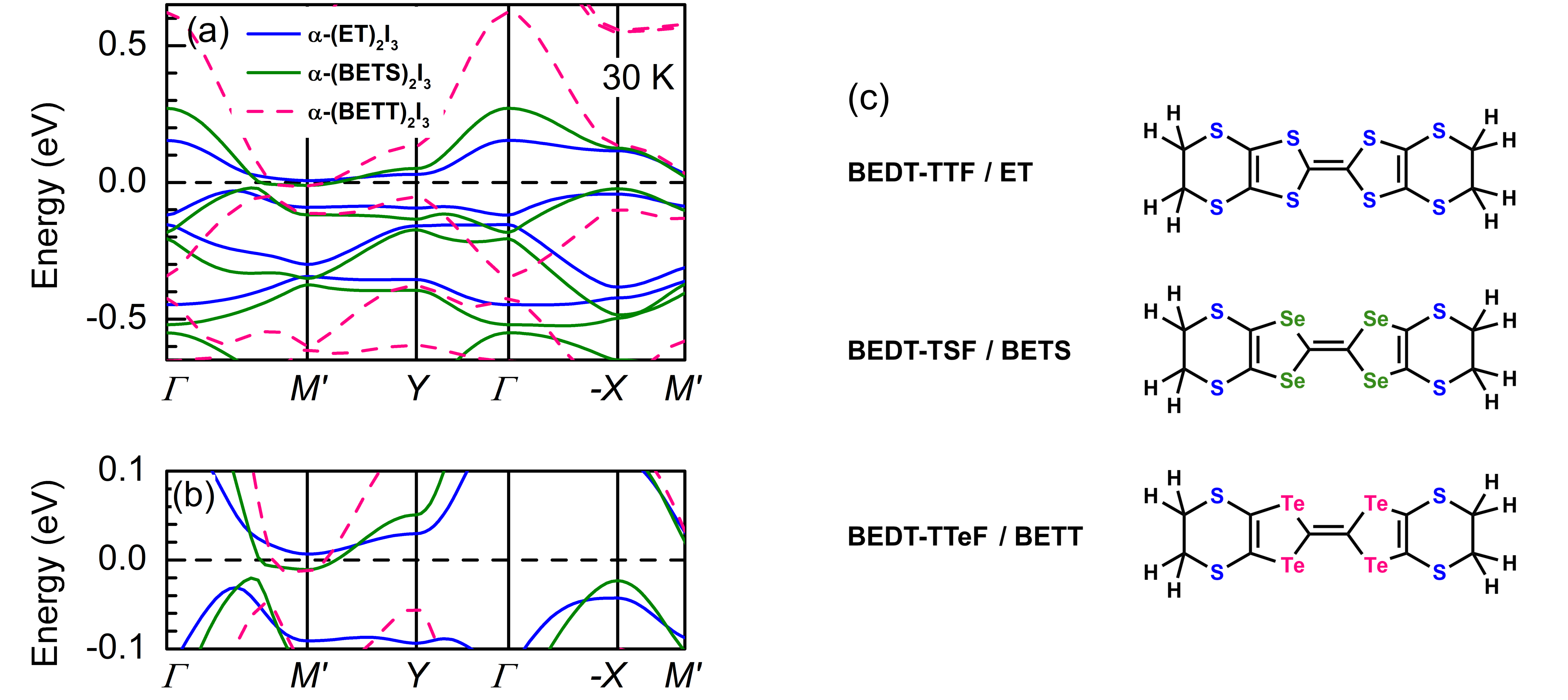}
        \caption{ (a) Band structure of \ETI, \BETSI{} and $\alpha$-(BETT)$_2$I$_3$ for $T=30$~K, (b) magnified plot of panel (a) near $E_F$. (c) The chemical structures of the three constituting molecules. }
    \label{fig:FigS7}
\end{figure}

To further study the role of chemical pressure on the electronic structure, we replace Se atoms in BEDT-TSF by Te atoms, leading to BEDT-TTeF (abbreviated for bis\-(ethylene\-dithio)\-tetra\-tellurium\-fulvalene). Although it has not been actually synthesized, we theoretically study BEDT-TTeF, where the inner S atoms are substituted by Te atoms (we utilize the crystal structure parameters for \BETSI). As illustrated in Fig.~\ref{fig:FigS7}, the low-temperature bandwidth increases when going from S via Se to Te atoms, but the gap between the two bands near Fermi level remains largely unchanged. For \ETI,  $E_F$ falls  in the aforementioned gap, making the ground state insulating, while for \BETSI{} and $\alpha$-(BETT)$_2$I$_3$, the gap lies just below the $E_F$, making the ground state metallic. These results elucidate the tunability of the bandwidth by chemical pressure mediated by the inner elements in the ET molecules, calling for further studies and synthesis of novel organic molecular conductors.

\newpage

\section{Intraband transition: Localized carriers}

\begin{figure}
    \centering
        \includegraphics[width=0.75\linewidth]{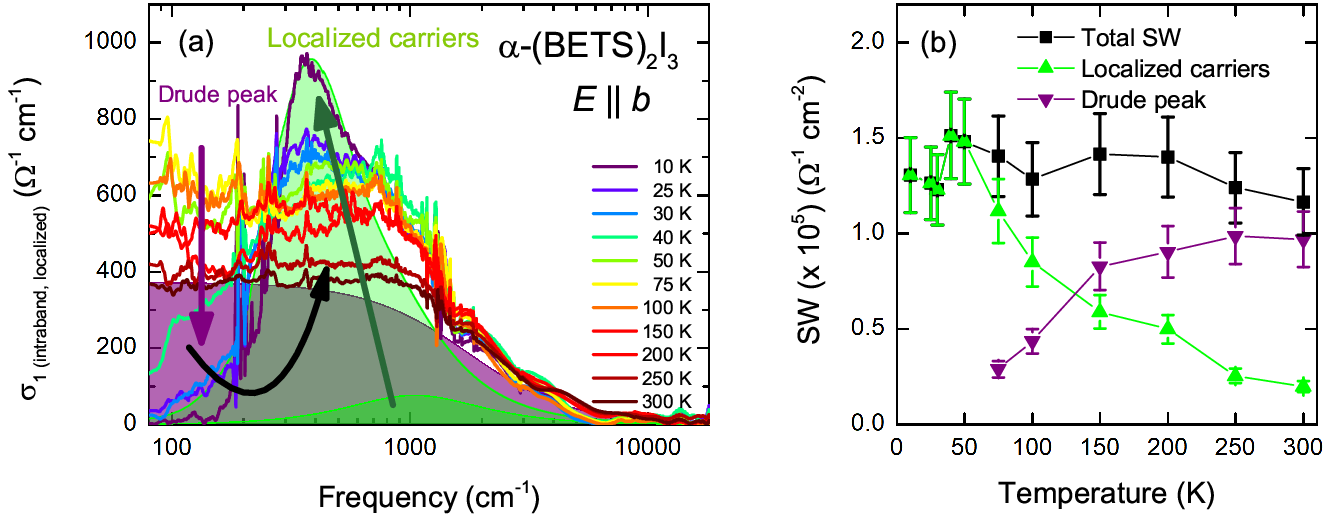}
        \caption{(a)~Optical conductivity (interband transitions and molecular vibrations are subtracted) of \BETSI{} for different temperatures. The Drude term is highlighted by violet, while the localization peak is indicated by green. (b) Temperature-dependent shift of the spectral weight (SW) from the Drude peak to the localized peak.}
    \label{fig:FigS8}
\end{figure}

\begin{figure}
    \centering
        \includegraphics[width=0.75\linewidth]{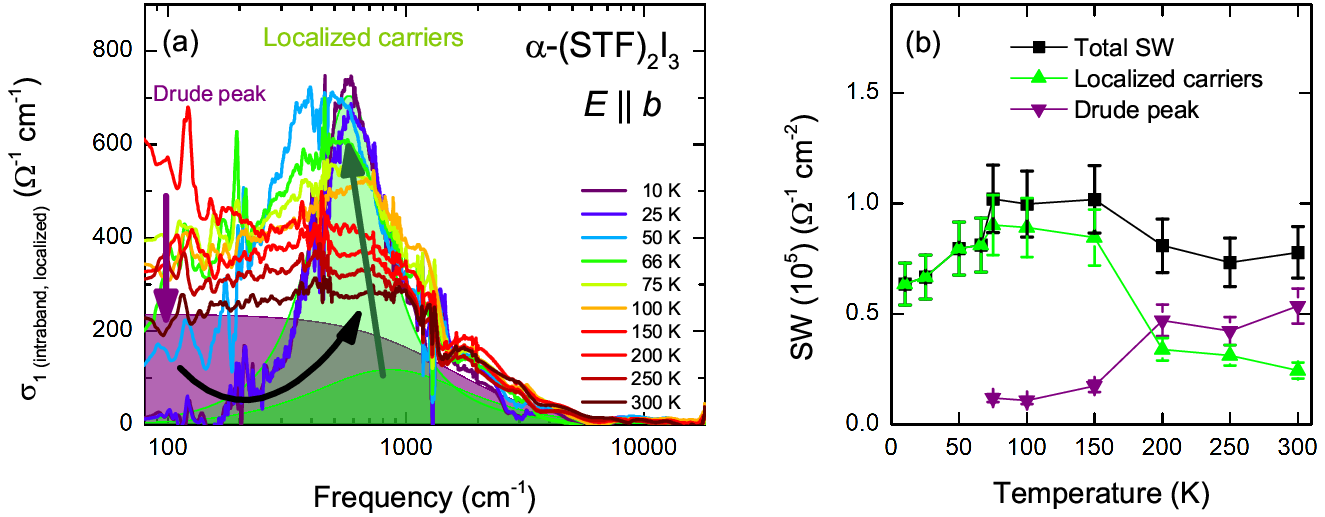}
        \caption{(a)~Temperature evolution of the optical conductivity (interband transitions and molecular vibrations are subtracted) for \STFI{}. The Drude term is highlighted by violet, while the localization peak is indicated by green. (b)~Shift of spectral weight (SW) from Drude peak to localized peak as a function of temperature.}
    \label{fig:FigS9}
\end{figure}

For \BETSI{} and \STFI, the interband transitions show little to no changes and the prominent alterations in the optical response are driven by itinerant and localized carriers.
In order to illustrate these changes in the intraband transitions, Figs.~\ref{fig:FigS8} and \ref{fig:FigS9} display the optical conductivity $\sigma_1(\omega)$  after the interband transitions and molecular vibrations have been subtracted. As the temperature is reduced, the overall behavior strongly deviates from a typical Drude behavior expected for conventional metallic systems. In Fig.~\ref{fig:FigS4} we have already seen that an additional finite-energy response is required to describe the intraband response of \STFI\ and \BETSI.
In the panels (a), we can see how the zero-energy response decreases upon cooling while the localization peak grows significantly and shifts to lower energies. This illustrates the gradual localization of the charge carriers in the insulator phase.
In the panels (b), the temperature dependence of the respective spectral weight is plotted.
While the total spectral weight remains conserved down to lowest temperature, the Drude-like contributions decrease and eventually vanish. The charge localization becomes more prominent with decreasing temperature, marked by the shift of spectral weight, resulting in the insulating state at low temperatures .

\newpage
\section{Low-energy optical response}

\begin{figure}
    \centering
        \includegraphics[width=0.45\linewidth]{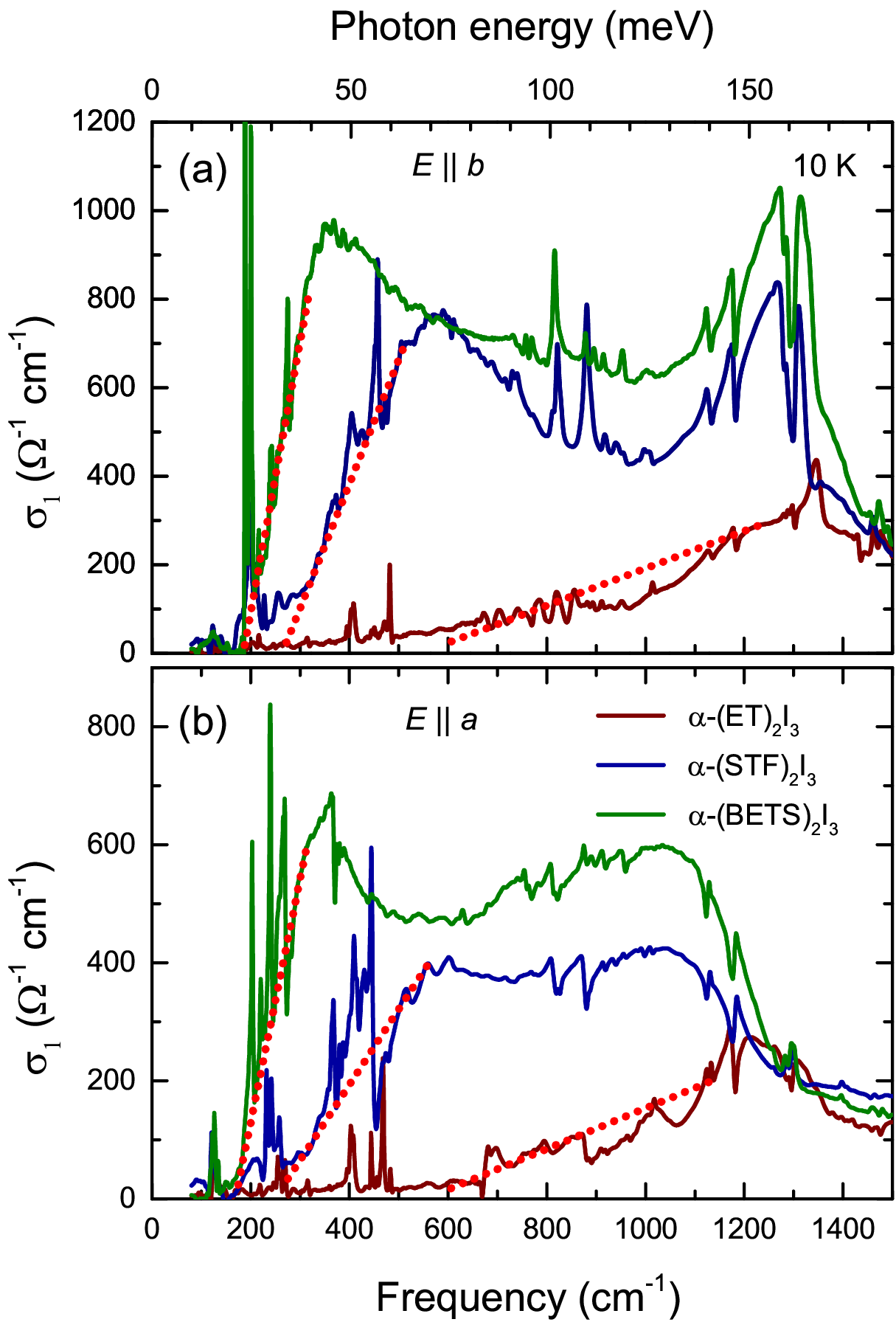}
        \caption{Low energy optical response at $T=10$~K for (a) $E \parallel b$ and (b) $E \parallel a$. The linear extrapolation of the slope in the region is utilized to estimate the energy gap, $\Delta E_{\rm gap}$ in the insulating phase for \ETI, \STFI, and \BETSI.}
    \label{fig:FigS10}
\end{figure}

The low-energy response is vital and reflects the driving force of the phase transition, which is the gradual localization leading to the opening of the energy gap at $T_{\rm MIT}$  as illustrated in Fig.~\ref{fig:FigS10}). Here, the low-energy optical conductivity of the insulating state is nearly extrapolated to low energies, with the x-axis crossing marking the gap energy of the insulating state as reported in Table~I of the main text.

\makeatletter
\renewcommand{\@biblabel}[1]{[S#1]}
\makeatother
\bibliography{ref}